\documentclass[manuscript, acmsmall]{acmart}
\newcommand{\commentout}[1]{}
\usepackage{soul}
\usepackage{enumerate}
\usepackage{enumitem}
\usepackage{multirow}
\usepackage{booktabs}
\usepackage{graphicx}
\usepackage{caption}
\usepackage{subcaption}
\usepackage{tabularx}
\usepackage{colortbl}
\usepackage{highlight}
\usepackage{hhline}
\usepackage{soul}
\usepackage{tabularray}
\usepackage[symbol]{footmisc}

\definecolor{pumpkin}{RGB}{211, 84, 0}
\newcommand{\ning}[1]{{\textcolor{black}{#1}}}

\AtBeginDocument{%
  \providecommand\BibTeX{{%
    \normalfont B\kern-0.5em{\scshape i\kern-0.25em b}\kern-0.8em\TeX}}}

\copyrightyear{2023} 
\acmYear{2023} 
\setcopyright{acmlicensed}
\acmDOI{}
\acmISBN{}

\begin{document}

\title[]{Insights into Natural Language Database Query Errors: From Attention Misalignment to User Handling Strategies}



\author{Zheng Ning}
\authornote{These authors contributed equally to this work.}
\affiliation{%
  \institution{University of Notre Dame}
  \city{Notre Dame}
  \state{IN}
  \country{USA}}
\email{zning@nd.edu}

\author{Yuan Tian}
\authornotemark[1]
\affiliation{%
  \institution{Purdue University}
  \city{West Lafayette}
  \state{IN}
  \country{USA}}
\email{tian211@purdue.edu}

\author{Zheng Zhang}
\affiliation{%
  \institution{University of Notre Dame}
  \city{Notre Dame}
  \state{IN}
  \country{USA}}
\email{zzhang37@nd.edu}

\author{Tianyi Zhang}
\affiliation{%
  \institution{Purdue University}
  \city{West Lafayette}
  \state{IN}
  \country{USA}}
\email{tianyi@purdue.edu}

\author{Toby Jia-Jun Li}
\affiliation{%
  \institution{University of Notre Dame}
  \city{Notre Dame}
  \state{IN}
  \country{USA}}
\email{toby.j.li@nd.edu}

\begin{abstract}

    Querying structured databases with natural language (NL2SQL) has remained a difficult problem for years. Recently, the advancement of machine learning (ML), natural language processing (NLP), and large language models (LLM) have led to significant improvements in performance, with the best model achieving $\sim$85\% percent accuracy on the benchmark Spider dataset. However, there is a lack of a systematic understanding of the types, causes, and effectiveness of error-handling mechanisms of errors for erroneous queries nowadays. To bridge the gap, a taxonomy of errors made by four representative NL2SQL models was built in this work, along with an in-depth analysis of the errors. Second, the causes of model errors were explored by analyzing the model-human attention alignment to the natural language query. Last, a within-subjects user study with 26 participants was conducted to investigate the effectiveness of three interactive error-handling mechanisms in NL2SQL. Findings from this paper shed light on the design of model structure and error discovery and repair strategies for natural language data query interfaces in the future.

\end{abstract}



\maketitle

\section{Introduction}
Data querying is an indispensable step in data analysis, sensemaking, and decision-making processes\footnote[7]{\ning{This work extends our previous research \cite{iui_paper} published at IUI with new experiments on large language models (GPT-4) and a root cause analysis of the model attention and human attention.}}. However, traditional data query interfaces require users to specify their queries in a formal language such as SQL, leading to significant learning barriers for non-expert users who have little programming experience~\cite{Shuaily2016AFF, Schlager1986ACM}. This problem becomes increasingly important in the Big Data era, given the rising needs for end users in many key domains including business, healthcare, public policy, scientific research, etc. To address this problem, natural language (NL) data query interfaces allow users to express data queries in natural language. For example, a semantic parser can map the user's natural language query into a formal data query language such as SQL (NL2SQL). Such NL interfaces have shown the potential to lower the bar for data querying and support flexible data exploration for end users~\cite{fu2023catsql,wang2023nalmo,baik2019bridging,tablan2008natural}.

However, achieving robust NL2SQL parsing in realistic scenarios is difficult because of the ambiguity in natural language and the complex structures (e.g., nested queries, joined queries) in the target queries. For example, in Spider~\cite{yu-etal-2018-spider}, a large-scale complex and cross-domain dataset for NL2SQL parsing, the accuracy of state-of-the-art models remained low in the 20\% to 30\% range for quite some time until 2019. In the past three years, advances in deep learning and large language models have brought us closer than ever to achieving useful performance on this important task---with the use of state-of-the-art end-to-end models such as~\cite{gao2023texttosql,pourreza2023dinsql,gan2021natural,huang2021relation}, the accuracy quickly increased to about 85\%. However, the development in model performance appears to have stagnated in the 85\% range recently, suggesting a bottleneck in model-only methods for NL2SQL.

This work focuses on the flip side of the 85\% accuracy---the 15\% erroneous queries. We started by understanding \textit{``What errors current NL2SQL models make.''} Then, we investigate \textit{``How NL2SQL models made these errors''} and \textit{``How users handle these errors''} with different studies.

We first performed a comprehensive analysis of SQL errors made by representative state-of-the-art NL2SQL models and developed an axial taxonomy of those errors. We reproduced four representative high-performing models with various structures from the Spider leaderboard\footnote{\url{https://yale-lily.github.io/spider}}---DIN-SQL + GPT-4~\cite{pourreza2023din}, SmBop + GraPPa (SmBop)~\cite{rubin2020smbop}, BRIDGE v2 + BERT (BRIDGE)~\cite{lin2020bridging}, and GAZP + BERT (GAZP)~\cite{zhong-etal-2020-grounded}. For each model, we collected all model-generated queries for the Spider dataset whose execution results varied from the ground truth results. Four authors conducted multiple rounds of qualitative coding and refinement on these errors to derive a taxonomy of the errors. The error analysis reveals that, despite having different model architectures and performance, NL2SQL errors originate from a common set of queries and demonstrate a similar distribution across various error types. \looseness=-1


Given the error distribution, we further investigate the potential reasons behind the error.
Inspired by recent work~\cite{attention1, attention6, attention3}, which showed that attention alignment can improve the performance in code summarization, machine translation, and visual feature representation, we hypothesize that SQL errors generated by NL2SQL models derive from the misalignment between the model's attention and human's attention toward the natural language query. To validate this hypothesis, two authors (SQL experts) manually annotated important words (\textit{human attention}) in the NL queries when they try to understand the query. The model-focused words (\textit{model attention}) were obtained by calculating the weight of each word that contributed to the model's prediction using a perturbation-based method~\cite{10.1109/TVCG.2018.2865230}. The attention alignment was measured by computing the overlap between the human-focused words and the model-focused words. The results showed a significant difference in attention alignment between erroneous queries and correctly predicted queries, implying that NL2SQL errors are highly correlated with attention misalignment. The findings suggest the promise of future work in aligning model attention with human attention to improve model performance.

To support NL2SQL models in real-world deployment, it is also important to provide effective error-handling mechanisms for users, which enable human users to discover and repair errors. In both human-computer interaction (HCI) and natural language processing (NLP) communities, we have seen efforts using different approaches. In general, there are three representative paradigms. First, following the \textit{task decomposition} paradigm, strategies such as DIY~\cite{narechania2021diy} decompose a generated SQL statement into a sequence of incremental subqueries and provide step-by-step explanations to help users identify errors. Second, based on query visualization, approaches such as QueryVis~\cite{leventidis2020queryvis}, QUEST~\cite{bergamaschi2013quest}, and SQLVis~\cite{SQLVis} seek to improve user understanding of the generated SQL statements by visualizing the structures and relations among entities and tables in a query. Third, using conversational agents, works such as ~\cite{misp,wang2021interactive,gur2018dialsql} implement chatbots to communicate the model's current state with users and update the results with user feedback through dialogs.
These interactive approaches help the model and humans to synchronize their attention. Consequently, humans gain confidence in the decisions generated by the model, while the models can make better decisions under human guidance.

Although these approaches were shown to be useful in different contexts in individual evaluations, it is unclear how effective each approach is for users with various SQL expertise. Furthermore, as most of these methods were evaluated with only simple NL2SQL errors, it is unclear how well they will perform on errors made by state-of-the-art NL2SQL models on complex datasets such as Spider. Therefore, we selected three representative models and conducted a controlled user study ($N=26$) to investigate the effectiveness and efficiency of representative error discovery and repair approaches. Specifically, we selected (i) an explanation- and example-based approach that supports fixing the SQL query through entity mapping between the natural language (NL) question and the generated query and discovering the error through a step-by-step NL explanation approach (DIY~\cite{narechania2021diy}), (ii) an explanation-based SQL visualization approach (SQLVis~\cite{SQLVis}), and (iii) a conversational dialog approach~\cite{misp}. The study reveals that these error-handling mechanisms have limited impacts on increasing the efficiency and accuracy of error discovery and repair for errors made by state-of-the-art NL2SQL models. Finally, we discussed the implications for future error-handling mechanisms in NL query interfaces.

To conclude, this paper presents the following four contributions:
\begin{itemize}[nosep]
    \item We developed a taxonomy of error types for three representative state-of-the-art NL2SQL models through iterative and axial coding procedures.
    \item We conducted a comprehensive analysis that compared model attention to human attention. The result shows that NL2SQL errors are highly correlated with attention misalignment between humans and models.
    \item We conducted a controlled user study that investigated the effectiveness and efficiency of three representative NL2SQL error discovery and repair methods.
    \item We discussed the implications for designing future error-handling mechanisms in natural language query interfaces. \looseness=-1
\end{itemize}
\section{Related Work}
    
    \subsection{NL2SQL techniques}
    
    Supporting natural language queries for relational databases is a long-standing problem in both the database (DB) and NLP communities. Given a relational database $D$ and a natural language query $q_{nl}$ to $D$, an NL2SQL model aims to find an equivalent SQL statement $q_{sql}$ to answer $q_{nl}$. The early methods of mapping $q_{nl}$ to $q_{sql}$ depend mainly on the development of intermediate logical representation \cite{warren-pereira-1982-efficient, grosz1983team} or mapping rules \cite{popescu2004modern, li2014constructing, saha2016athena, yaghmazadeh2017sqlizer, baik2019bridging}. In the former case, $q_{nl}$ is first parsed into logical queries independent of the underlying database schema, which are then converted into queries that can be executed on the target database \cite{kim2020natural}. On the contrary, rule-based methods generally assume that there is a one-to-one correspondence between the words in $q_{nl}$ and a subset of database keywords/entities \cite{kim2020natural}. Therefore, the NL2SQL mapping can be achieved by directly applying the syntactic parsing and semantic entity mapping rules to $q_{nl}$. Although both strategies have achieved significant improvement over time, they have two intrinsic limitations. First, they require significant effort to create hand-crafted mapping rules for translation \cite{kim2020natural}. Second, the coverage of these methods is limited to a definite set of semantically tractable natural language queries \cite{popescu2004modern, kim2020natural}. 
    
    The recent development of deep learning (DL) based methods aims to achieve flexible NL2SQL translation through a data-driven approach \cite{rubin2020smbop, lin2020bridging, zhong-etal-2020-grounded, iyer2017learning, zhong2017seq2sql, bogin2019representing, guo2019towards}. From large-scale datasets, DL-based models learn to interpret NL queries conditioned on a relational DB via SQL logic~\cite{lin2020bridging}. Most NL2SQL models use the encoder-decoder architecture~\cite{lin2020bridging, zhong2017seq2sql, zhong-etal-2020-grounded}, where the encoder models the input $q_{nl}$ into a sequence of hidden representations along time steps. The decoder then maps the hidden representations into the corresponding SQL statement. Recently, Transformer-based architecture \cite{wang2019rat, lin2020bridging} and pre-training techniques \cite{yin2020tabert, herzig2020tapas, scholak-etal-2021-picard} have become popular as the backbone of NL2SQL encoders. At the same time, many decoders have been used to optimize SQL generation, such as autoregressive bottom-up decoding~\cite{rubin2020smbop} and the LSTM-based pointer-generator \cite{lin2020bridging}. However, those DL-based models are usually ``black-boxes'' due to the lack of explainability~\cite{kim2020natural}. The lack of transparency makes it difficult for users to figure out how to fix the observed errors when using DL-based NL2SQL models. 
    
    The evaluation of these DL-based models is based mainly on objective benchmarks such as Spider~\cite{yu-etal-2018-spider} and WikiSQL~\cite{zhong2017seq2sql}. For example, Spider requires models to generalize well not only to unseen NL queries but also to new database schemas, in order to encourage NL interfaces to adapt to cross-domain databases. The performance of a model is evaluated using multiple measures, including component matching, exact matching, and execution accuracy. However, these benchmarks only involve quantitative analysis of NL2SQL models, giving little clue about what types of errors a model tends to fall into.
    
    An aim of our work is to develop a taxonomy of error types of errors made by state-of-the-art NL2SQL models and report the corresponding descriptive statistics to complement the quantitative benchmark with the qualitative analysis of those NL2SQL models.
    
    \subsection{Detecting and repairing errors for NL2SQL}
    \label{sec:error_detect_repair}
    
    Natural language interfaces for NL2SQL face challenges in language ambiguity, underspecification, and model misunderstanding \cite{narechania2021diy, elgohary2020speak}. Previous work has explored ways to support error detection and repair for NL2SQL systems through human-AI collaboration. NL2SQL error detection methods can be mainly divided into categories of natural language-explanation-based, visualization-based, and conversation-based approaches. NL2SQL error repair methods consist mainly of direct manipulation and conversational error fixing approaches.
    
    For error detection, a popular method is to explain the query and its answer in natural language \cite{elgohary2020speak, misp, simitsis2009dbmss, su2017building, tian-etal-2023-interactive}. For example, DIY \cite{narechania2021diy} and STEPS \cite{tian-etal-2023-interactive} show step-by-step NL explanations and query results by applying templates to subqueries, helping users understand SQL output in an incremental way; similarly, SOVITE~\cite{Li2020MultiModalRO} allows users to fix agent breakdowns in understanding user instructions by revealing the agent's current state of understanding of the user's intent and supporting direct manipulation for the user to repair the detected errors; NaLIR \cite{10.1145/2588555.2594519} explains the mapping relations between entities in the input query and those in the database schema; Ioannidis et al. \cite{simitsis2009dbmss} introduced a technique to describe structured database content and SQL translation textually to support user sensemaking of the model output. Visualizations have also been widely used to explain a SQL query and its execution \cite{berant2019explaining, bergamaschi2013quest, SQLVis, leventidis2020queryvis}. For example, QueryVis~\cite{leventidis2020queryvis} produces diagrams of SQL queries to capture their logical structures; QUEST~\cite{bergamaschi2013quest} visualizes the connection between the matching entities from the input query and their correspondences in the database schema; SQLVis \cite{SQLVis} introduced visual query representations to help SQL users understand the complex structure of SQL queries and verify their correctness.
    
    Most of the previous work employed direct manipulation to repair and disambiguate queries. NaLIR~\cite{10.1145/2588555.2594519}, DIY~\cite{narechania2021diy} and DataTone~\cite{gao2015datatone} allow users to modify the entity mappings through drop-downs; Eviza~\cite{setlur2016eviza} and Orko~\cite{srinivasan2017orko} enable users to modify quantitative values in queries through range sliders. In addition to direct manipulation, several other prior interaction mechanisms enable users to give feedback to NL2SQL models through dialogs in natural language. For example, MISP~\cite{misp} maintains the state of the current parsing process and asks for human feedback to improve SQL translation through human-AI conversations. Elgohary et al.~\cite{elgohary2020speak, Elgohary2021NLEDITCS} investigated how to enable users to correct  NL2SQL parsing results with natural language feedback in conversation. 
    
    With the many error-handling mechanisms that have been proposed, there is a gap in evaluating how effective and efficient these mechanisms are to address different types of NL2SQL errors and what specific limitations they have. These types of information are critical to inform the effective choice and design of NL2SQL error handling mechanisms in different use scenarios and to inspire the use of ensemble mechanisms to handle different usage contexts of NL2SQL. Our work bridges this gap by investigating these questions through controlled user studies, whose findings could guide the future design of NL2SQL error handling systems.

    \subsection{Error handling via human-AI collaboration}
    
    Handling errors made by AI models in human-AI collaboration faces many key challenges. First, many state-of-the-art AI models lack transparency in their decision-making process, making it difficult for users to understand exactly what leads to incorrect predictions \cite{samek2017explainable}. Although there are some attempts to explain the state of the AI model using methods such as heatmap \cite{rieger2020simple, zhou2018interpretable}, search traces~\cite{zhang2021interpretable}, and natural language explanations \cite{ehsan2018rationalization, costa2018automatic}, they only allow users to peek at the AI model's reasoning at certain stages instead of exposing the holistic states of the model. Second, it is difficult for users to develop a correct mental model for complex AI models due to the \textit{representational mismatch} in which ``{\em humans can create a mental model in terms of features that are not identical to those used by AI models}''~\cite{bansal2019beyond}. Lastly, error handling usually requires multiple turns of interactions~\cite{Li2020MultiModalRO, lai2022type}. However, maintaining coherent multi-turn interactions between AI and humans is challenging \cite{aliannejadi2020harnessing}. It requires AI to closely maintain and update the context history, evolve its contextual understanding, and behave appropriately based on user's timely responses~\cite{zhou2018multi, allen2007plow, 10.3115/981863.981872}.
    
    Our work contributes to the knowledge of how users handle errors in their collaborations with NL2SQL models by studying how users utilize existing error-handling mechanisms to inspect and fix errors made by NL2SQL models and how they perceive the usefulness of these mechanisms. Our findings of user challenges also echo the identified challenges in human-AI collaborations in other domains (e.g., programming~\cite{vaithilingam2022expectation, tang2023empirical, zhang2020interactive}, data annotation~\cite{gebreegziabher_patat:_2023}, QA generation~\cite{zhang_storybuddy_2022}, interactive task learning~\cite{li_sugilite:_2017, li_pumice:_2019}), showing that users need help comprehending the state of AI models and developing a proper mental model in AI-based interactive data-centric tools to understand and assess their recommendations.
 
\section{An Analysis and Taxonomy of NL2SQL Errors}
\label{sec:analysis}

In this section, we describe the development of the taxonomy of NL2SQL errors of four representative NL2SQL models and the corresponding error analysis. The structure of this section is as follows. Section~\ref{sec:model_selection} introduces the models we selected for analyzing the erroneous SQL queries. We also discussed the discrepancy in the  workflow of different models. Section~\ref{sec:data_collection} summarizes the methodology used for building the dataset; Section~\ref{sec:coding} explains the axial and iterative coding procedure we used to derive the error taxonomy; Section~\ref{sec:taxonomy} describes the developed taxonomy of NL2SQL errors; Section~\ref{sec:error_analysis} presents an analysis of erroneous queries in the dataset based on the taxonomy.

\subsection{Model selection}
\label{sec:model_selection}
We selected four representative NL2SQL models from the official Spider leader board, the information of which is shown in Table~\ref{tab: retrain-stats}.

While most models generate the SQL query in a top-down decoding procedure, SmBop (M1) improves the speed and accuracy by adopting a bottom-up structure. Specifically, it gradually combines smaller components to form a syntactically larger SQL component.
BRIDGE (M2) is a sequential architecture that models the dependencies between natural language queries and relational databases. It combines a BERT-based~\cite{devlin_2018_bert} encoder with a sequential pointer-generator for end-to-end cross-DB NL2SQL semantic parsing. In comparison, GAZP (M3) combines a semantic parser with an utterance generator. When given a new database, it synthesizes data and selects instances that are consistent with the schema to adapt an existing semantic parser to this new database.
DIN-SQL+GPT-4 (M4) has the best accuracy among all four models. It improves the performance of the large language model (GPT-4~\cite{openai2023gpt4}) on text-to-SQL tasks by employing task decomposition, adaptive prompting, linking schema to prompt, and self-correction.
All these models employ different NL2SQL task-solving strategies at different stages, including decoding (M1), encoding (M2), finetuning (M3), and task preprocessing (M4).

\begin{table}
\centering
 \scalebox{0.79}{
\begin{tabular}{|c|c|c|c|c|} 
\hline
\textbf{Model Index} & \textbf{Model Names} & \textbf{Err. queries} & \textbf{Retrained Acc.} & \textbf{Original Acc.} \\
\hline
M1 & SmBop~\cite{rubin2020smbop}  & 431                               & 81.2\%                      & 71.1\%                                                          \\ 
\hline
M2 & BRIDGE~\cite{lin2020bridging} & 853                               & 62.7\%                      & 68.3\%                                                          \\ 
\hline
M3 & GAZP~\cite{zhong2020grounded}   & 1062                              & 53.6\%                      & 53.5\%                                                          \\
\hline
M4 & DIN-SQL+GPT-4~\cite{pourreza2023din} & 304                              & 86.7\%                      & 85.3\%                                                          \\
\hline
\end{tabular}
}
\caption{Descriptive statistics and the accuracy of each model we reproduced}
\label{tab: retrain-stats}
\end{table}

\subsection{Erroneous queries dataset collection}
\label{sec:data_collection}
We adopted the Spider~\cite{yu-etal-2018-spider} dataset to train and evaluate the models and to collect a set of erroneous SQL queries for the taxonomy. Spider is the most popular benchmark to evaluate NL2SQL models with complex and cross-domain semantic parsing problems. It consists of around 10,000 queries in natural language on multiple databases across different domains (e.g., ``soccer'', ``college''). In the original Spider dataset, the difficulty of the queries is divided into four levels: ``Easy'', ``Medium'', ``Hard'', and ``Extra Hard'', depending on the complexity of their structures and the SQL keywords involved. We demonstrate an example NL-SQL pair for each difficulty level in Table~\ref{tab:difficulty_examples}. In this work, we focus only on the first three difficulty levels as state-of-the-art models have significantly lower accuracy on the ``extra hard'' queries --- the best model we reproduced, DIN-SQL+GPT-4, only achieved less than 50\% in accuracy, indicating that NL2SQL for ``extra hard'' queries remains less feasible at this point.




\begin{table*}
\centering

\captionsetup{labelformat=empty}
\resizebox{0.94\textwidth}{!}{
\begin{tblr}{
  column{1} = {c},
  hlines,
  vline{2-3} = {-}{},
}
                                & \textbf{NL query}                                                                                                                                       & \textbf{SQL query}                                                                                                                                                                                                                                                                                                                                                      \\
\textbf{Easy}                   & {What is the abbreviation \\for airline \`{}\`{}JetBlue Airways'' ?}                                                                                    & {SELECT Abbreviation FROM AIRLINES \\WHERE Airline = \`{}\`{}JetBlue Airways''}                                                                                                                                                                                                                                                                                         \\
\textbf{Medium}                 & {What are the codes of countries \\where Spanish is spoken by the \\largest percentage of people?}                                                      & {SELECT CountryCode , MAX(Percentage) FROM \\countrylanguage WHERE language= \`{}\`{}Spanish'' \\GROUP BY CountryCode}                                                                                                                                                                                                                                                  \\
\textbf{Hard}                   & {What are the first names of the students \\who live in Haiti permanently or \\have the cell phone number 09700166582?}                                 & {SELECT T1.first\_name FROM students AS T1 \\JOIN addresses AS t2 ON T1.permanent\_address\_id = T2.address\_id \\WHERE T2.country = 'haiti' OR T1.cell\_mobile\_number = '09700166582’}                                                                                                                                                                                \\
{\textbf{Extra}\\\textbf{Hard}} & {What is the series name and country of \\all TV channels that  are playing cartoons \\directed by Ben Jones and cartoons \\directed by Michael Chang?} & {SELECT T1.series\_name , T1.country FROM TV\_Channel AS T1 \\JOIN cartoon AS T2 ON T1.id = T2.Channel WHERE T2.directed\_by = \\'Michael Chang' INTERSECT SELECT T1.series\_name , T1.country \\FROM TV\_Channel AS T1 JOIN cartoon AS T2 ON T1.id = T2.Channel \\WHERE T2.directed\_by = 'Ben Jones'} 

\end{tblr}
}
\caption{Table 2. \ NL-SQL pairs with different difficulty levels in the Spider dataset}
\label{tab:difficulty_examples}
\end{table*}

Since the held-out test set in Spider is not publicly available, we created our own test set by re-splitting the public training and development sets from Spider. The ratios of the three difficulty levels in the new training and testing sets were close to those in the original training and developing sets. In addition, we ensured that there was no overlap in the databases used between our training and testing sets. Table~\ref{tab: split} shows the distribution of our training and test data compared to the original public Spider dataset. We re-trained model M1-M3  with their officially released code using our training set. Since M4 was using few-shot learning, we did not retrain this model separately. The models used different core structures and showed close to SOTA performance at the time we conducted the study. \looseness=-1

The erroneous queries are those that have different execution results from the correct ones (with values). The queries generated by these models on the test set were manually analyzed to develop the taxonomy of SQL generation errors. Table~\ref{tab: retrain-stats} shows the total number of erroneous queries generated by each model. The accuracy of each model on our test set is close to the reported performance of these models on the private held-out test set, indicating that our reproduction of these models are consistent with the original implementations.

\begin{table}
\centering
 \scalebox{0.83}{
\begin{tabular}{|c|c|c|c|c|c|c|} 
\hline
                                          &       & \textbf{Easy} & \textbf{Medium} & \textbf{Hard} & \textbf{Extra} & \textbf{Total}  \\ 
\hline
\multirow{2}{*}{\textbf{Original}} & Train & 1983          & 2999            & 1921          & 1755                & 8658            \\ 
\cline{2-7}
                                          & Dev   & 248           & 446             & 174           & 166                 & 1034            \\ 
\hline
\multirow{2}{*}{\textbf{Re-split}} & Train & 1604          & 2363            & 1516          & 0                   & 5483            \\ 
\cline{2-7}
                                          & Test  & 627           & 1082            & 579           & 0                   & 2288            \\
\hline
\end{tabular}
}
\caption{Descriptive statistics of the original Spider dataset and our sampled dataset}
\label{tab: split}
\end{table}

\subsection{The coding procedure}
\label{sec:coding}

After curating the dataset of erroneous queries, we followed the established open, axial and iterative coding process~\cite{antoine2021interaction, lazar2017research,braun2006using} to develop a taxonomy of NL2SQL errors. The detail of the process is as follows.

\subsubsection{Step 1: Open coding}
To begin with, we randomly sampled 40 erroneous SQL queries to develop the preliminary taxonomy. Four authors with in-depth SQL knowledge performed open coding~\cite{antoine2021interaction,lazar2017research,braun2006using} on this subset of erroneous SQL queries. They were instructed to code to answer the following questions: (1) \textit{What are the errors in the generated SQL query in comparison to the ground truth?} (2) \textit{What SQL component does each error reside at?}  (3) \textit{Have all the errors in the incorrect SQL query been covered?}. Once finishing the first round of coding, the coded query pairs (the generated query and the ground truth) were put line by line in a shared spreadsheet. The annotators sat together to discuss the codes and reached a consensus on the preliminary version of the codebook. 

\subsubsection{Step 2: Iterative refinement of the codebook}
\label{sec:refine}
After creating the preliminary codebook, four annotators conducted iterative refinements of the established codes. Each iteration consisted of the following three steps. First, the annotators coded a new sample batch of 40 unlabeled erroneous queries using the codebook from the last iteration. If there is a new error not covered by the current codebook, annotators would write a short description of it. Second, we computed the inter-rater reliability between coders~\cite{10.1145/3359174} (Fleiss' Kappa and Krippendorff's Alpha) at the end of each iteration. Lastly, annotators exchanged opinions about adding, merging, removing, or reorganizing codes and updated the codebook accordingly. Annotators completed three refinement iterations until the codebook became stable and the inter-rater reliability scores were substantial. At the end of the final refinement iteration, the Fleiss' Kappa was 0.69 and the Krippendorff's Alpha was 0.67.

\subsubsection{Step 3: Coding the remaining dataset}
We then proceeded to code the remaining dataset using the codebook from the final refinement iteration. Because the inter-rater reliability scores stabilized among annotators, two annotators participated in this step. The Fleiss' Kappa and Krippendorff's Alpha of the full dataset annotation between those two annotators were 0.76 and 0.78 respectively, indicating substantial agreement~\cite{antoine2021interaction,fleiss1971measuring,krippendorff2011computing}.

\begin{figure*}
\centering
\includegraphics[width=\linewidth]{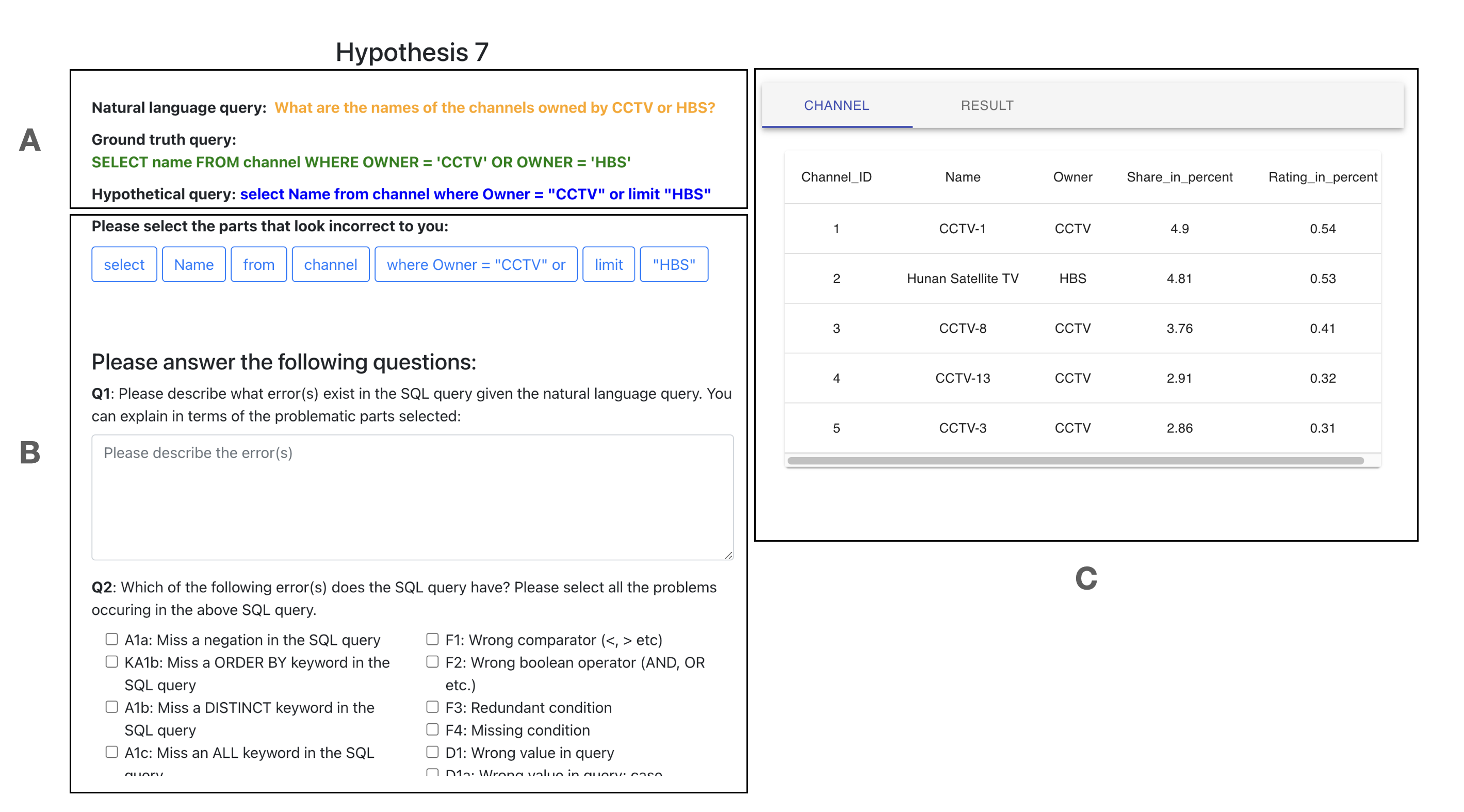}
\caption{The user interface that we used for NL2SQL error annotation}
\label{fig:f_annotation_ui}
\end{figure*}
    
\subsubsection{The annotation interface}

We implemented an error annotation system to annotate the erroneous queries. The front-end UI is shown in Figure \ref{fig:f_annotation_ui}. It consists of three components: (A): A query display section that presents the natural language query, the corresponding ground truth, and the model-generated SQL query. (B): An error annotation section where the annotator first decided which part(s) of the generated SQL is wrong by clicking on the corresponding selection button. After that, the annotator was supposed to choose error types from the checkbox below. If the error type was not included, the system provided an input box to accept open feedback. The error types would be updated after each batch of coding described in Section~\ref{sec:refine}. (C): A results display section. The tables involved in the pairs and the execution result were shown in this section to help annotators identify the error types.\looseness=-1

\subsection{The Taxonomy of NL2SQL Errors}
\label{sec:taxonomy}
Table~\ref{tab:taxonomy} shows the finalized taxonomy of NL2SQL errors. Specifically, we categorized the error types along two dimensions: (1) the \textit{syntactic} dimension shows which parts of the SQL query an error occurs in, categorized by SQL keywords such as \texttt{WHERE} and \texttt{JOIN}; (2) the \textit{semantic} dimension indicates which aspects of the NL description that the model misunderstands, such as misunderstanding a value or the name of a table. For each type of error, the uppercase letter refers to the syntactic category, and the lowercase letter refers to the semantic category. Note that there may be multiple manifestations of a semantic error in a syntactic error category. For example, the table error has two different forms in the \texttt{``JOIN''} clause, including \texttt{``Miss a table to JOIN''} (Ba1) and \texttt{``JOIN the wrong table''} (Ba2). An erroneous query may also have multiple error types associated with it.

\begin{table}[!ht]
\centering
\setlength{\extrarowheight}{0pt}
\addtolength{\extrarowheight}{\aboverulesep}
\addtolength{\extrarowheight}{\belowrulesep}
\setlength{\aboverulesep}{0pt}
\setlength{\belowrulesep}{0pt}
\renewcommand{\arraystretch}{0.7} 
\resizebox{0.98\linewidth}{!}{
\begin{tabular}{c|l|cl|cccc}
\toprule
Error categories                   & \multicolumn{3}{c|}{Error types}                                                    & SmBop                                   & BRIDGE                                  & GAZP                                      &\begin{tabular}[c]{@{}c@{}}DIN-SQL+\\GPT-4\end{tabular} \\ 
\hline
\multirow{40}{*}{Syntactic errors} & \multirow{10}{*}{A: WHERE error}   & Aa1: & Use a wrong table in WHERE              &{\cellcolor[rgb]{0.98, 0.92, 0.67}} 21   & {\cellcolor[rgb]{0.98, 0.92, 0.67}}68   &{\cellcolor[rgb]{0.98, 0.92, 0.67}} 73                                        &{\cellcolor[rgb]{0.98, 0.92, 0.67}}10 \\
                                   &                                    & Ab1: & Use a wrong column in WHERE             & {\cellcolor[rgb]{0.702,0.808,0.984}}19  & {\cellcolor[rgb]{0.702,0.808,0.984}}36  & {\cellcolor[rgb]{0.702,0.808,0.984}}23    &{\cellcolor[rgb]{0.702,0.808,0.984}}12\\
                                   &                                    & Ac1: & Redundant WHERE clause                  & {\cellcolor[rgb]{0.984,0.855,0.843}}14  & {\cellcolor[rgb]{0.984,0.855,0.843}}16  & {\cellcolor[rgb]{0.984,0.855,0.843}}27    &{\cellcolor[rgb]{0.984,0.855,0.843}}13\\
                                   &                                    & Ac2: & Missing WHERE clause                    & {\cellcolor[rgb]{0.984,0.855,0.843}}15  & {\cellcolor[rgb]{0.984,0.855,0.843}}21  & {\cellcolor[rgb]{0.984,0.855,0.843}}61    &{\cellcolor[rgb]{0.984,0.855,0.843}}7\\
                                   &                                    & Ad1: & Other wrong value in WHERE clause       & {\cellcolor[rgb]{0.82,0.945,0.855}}51   & {\cellcolor[rgb]{0.82,0.945,0.855}}52   & {\cellcolor[rgb]{0.82,0.945,0.855}}91     &{\cellcolor[rgb]{0.82,0.945,0.855}}18\\
                                   &                                    & Ad2: & Value case error in WHERE clause        & {\cellcolor[rgb]{0.82,0.945,0.855}}62   & {\cellcolor[rgb]{0.82,0.945,0.855}}69   & {\cellcolor[rgb]{0.82,0.945,0.855}}82     &{\cellcolor[rgb]{0.82,0.945,0.855}}36\\
                                   &                                    & Ad3: & Value plurality error in WHERE clause   & {\cellcolor[rgb]{0.82,0.945,0.855}}8    & {\cellcolor[rgb]{0.82,0.945,0.855}}6    & {\cellcolor[rgb]{0.82,0.945,0.855}}16     &{\cellcolor[rgb]{0.82,0.945,0.855}}0\\
                                   &                                    & Ad4: & Value synonym error in WHERE clause     & {\cellcolor[rgb]{0.82,0.945,0.855}}35   & {\cellcolor[rgb]{0.82,0.945,0.855}}40   & {\cellcolor[rgb]{0.82,0.945,0.855}}45     &{\cellcolor[rgb]{0.82,0.945,0.855}}35\\
                                   &                                    & Ae1: & Wrong comparator (<, >, =, !=, etc)       & {\cellcolor[rgb]{0.851,0.851,0.851}}8   & {\cellcolor[rgb]{0.851,0.851,0.851}}13  & {\cellcolor[rgb]{0.851,0.851,0.851}}14    &{\cellcolor[rgb]{0.851,0.851,0.851}}3\\
                                   &                                    & Ae2: & Wrong boolean operator (AND, OR etc.)   & {\cellcolor[rgb]{0.851,0.851,0.851}}4   & {\cellcolor[rgb]{0.851,0.851,0.851}}15  & {\cellcolor[rgb]{0.851,0.851,0.851}}9     &{\cellcolor[rgb]{0.851,0.851,0.851}}3\\ 
\hhline{~-------}
                                   & \multirow{4}{*}{B: JOIN error}     & Ba1: & Miss a table to JOIN                    &{\cellcolor[rgb]{0.98, 0.92, 0.67}} 35                                      &{\cellcolor[rgb]{0.98, 0.92, 0.67}} 106                                     &{\cellcolor[rgb]{0.98, 0.92, 0.67}} 101                                       &{\cellcolor[rgb]{0.98, 0.92, 0.67}}15 \\
                                   &                                    & Ba2: & JOIN the wrong table                    &{\cellcolor[rgb]{0.98, 0.92, 0.67}} 24                                      & {\cellcolor[rgb]{0.98, 0.92, 0.67}}89                                      & {\cellcolor[rgb]{0.98, 0.92, 0.67}}78                                        &{\cellcolor[rgb]{0.98, 0.92, 0.67}}3\\
                                   &                                    & Bb1: & Use a wrong column in JOIN              & {\cellcolor[rgb]{0.702,0.808,0.984}}13  & {\cellcolor[rgb]{0.702,0.808,0.984}}79  & {\cellcolor[rgb]{0.702,0.808,0.984}}69    &{\cellcolor[rgb]{0.702,0.808,0.984}}7\\
                                   &                                    & Bc1: & Redudant JOIN clause                    & {\cellcolor[rgb]{0.984,0.855,0.843}}17  & {\cellcolor[rgb]{0.984,0.855,0.843}}82  & {\cellcolor[rgb]{0.984,0.855,0.843}}113   &{\cellcolor[rgb]{0.984,0.855,0.843}}36\\ 
\hhline{~-------}
                                   & \multirow{4}{*}{C: ORDER BY error} & Cb1: & Use a wrong column to sort              & {\cellcolor[rgb]{0.702,0.808,0.984}}3   & {\cellcolor[rgb]{0.702,0.808,0.984}}26  & {\cellcolor[rgb]{0.702,0.808,0.984}}26    &{\cellcolor[rgb]{0.702,0.808,0.984}}9\\
                                   &                                    & Cc1: & Miss a ORDER BY clause                  & {\cellcolor[rgb]{0.984,0.855,0.843}}12  & {\cellcolor[rgb]{0.984,0.855,0.843}}22  & {\cellcolor[rgb]{0.984,0.855,0.843}}20    &{\cellcolor[rgb]{0.984,0.855,0.843}}4    \\
                                   &                                    & Cc2: & Redundant sorting                       & {\cellcolor[rgb]{0.984,0.855,0.843}}3   & {\cellcolor[rgb]{0.984,0.855,0.843}}1   & {\cellcolor[rgb]{0.984,0.855,0.843}}3     &{\cellcolor[rgb]{0.984,0.855,0.843}}0     \\
                                   &                                    & Ce1: & Wrong sorting direction                 & {\cellcolor[rgb]{0.851,0.851,0.851}}6   & {\cellcolor[rgb]{0.851,0.851,0.851}}27  & {\cellcolor[rgb]{0.851,0.851,0.851}}23    &{\cellcolor[rgb]{0.851,0.851,0.851}}15    \\ 
\cline{2-7}
                                   & \multirow{6}{*}{D: SELECT error}   & Da1: & Use a wrong table in SELECT             &{\cellcolor[rgb]{0.98, 0.92, 0.67}} 59                                      &{\cellcolor[rgb]{0.98, 0.92, 0.67}} 117                                     & {\cellcolor[rgb]{0.98, 0.92, 0.67}}106                                       &{\cellcolor[rgb]{0.98, 0.92, 0.67}}26                                       \\
                                   &                                    & Db1: & Return a wrong column in SELECT         & {\cellcolor[rgb]{0.702,0.808,0.984}}21  & {\cellcolor[rgb]{0.702,0.808,0.984}}56  & {\cellcolor[rgb]{0.702,0.808,0.984}}78    &{\cellcolor[rgb]{0.702,0.808,0.984}}33    \\
                                   &                                    & Db2: & Return a redundant column in SELECT     & {\cellcolor[rgb]{0.702,0.808,0.984}}10  & {\cellcolor[rgb]{0.702,0.808,0.984}}19  & {\cellcolor[rgb]{0.702,0.808,0.984}}36    &{\cellcolor[rgb]{0.702,0.808,0.984}}13    \\
                                   &                                    & Db3: & Miss returning column(s) in SELECT      & {\cellcolor[rgb]{0.702,0.808,0.984}}20  & {\cellcolor[rgb]{0.702,0.808,0.984}}34  & {\cellcolor[rgb]{0.702,0.808,0.984}}59    &{\cellcolor[rgb]{0.702,0.808,0.984}}11    \\
                                   &                                    & Df1: & Use wrong aggretation function          & {\cellcolor[rgb]{0.737,0.557,0.012}}7   & {\cellcolor[rgb]{0.737,0.557,0.012}}43  & {\cellcolor[rgb]{0.737,0.557,0.012}}11    &{\cellcolor[rgb]{0.737,0.557,0.012}}9    \\
                                   &                                    & Df2: & Miss aggregation function               & {\cellcolor[rgb]{0.737,0.557,0.012}}5   & {\cellcolor[rgb]{0.737,0.557,0.012}}19  & {\cellcolor[rgb]{0.737,0.557,0.012}}14    &{\cellcolor[rgb]{0.737,0.557,0.012}}6    \\ 
\hhline{~-------}
                                   & \multirow{3}{*}{E: GROUP BY error} & Eb1: & Use a wrong column in GROUP BY          & {\cellcolor[rgb]{0.702,0.808,0.984}}6   & {\cellcolor[rgb]{0.702,0.808,0.984}}10  & {\cellcolor[rgb]{0.702,0.808,0.984}}18    &{\cellcolor[rgb]{0.702,0.808,0.984}}12    \\
                                   &                                    & Ec1: & Miss a GROUP BY clause in the SQL query & {\cellcolor[rgb]{0.984,0.855,0.843}}10  & {\cellcolor[rgb]{0.984,0.855,0.843}}33  & {\cellcolor[rgb]{0.984,0.855,0.843}}47    &{\cellcolor[rgb]{0.984,0.855,0.843}}8    \\
                                   &                                    & Ec2: & Redudant GROUP BY clause                & {\cellcolor[rgb]{0.984,0.855,0.843}}6   & {\cellcolor[rgb]{0.984,0.855,0.843}}7   & {\cellcolor[rgb]{0.984,0.855,0.843}}16    &{\cellcolor[rgb]{0.984,0.855,0.843}}9    \\ 
\hhline{~-------}
                                   & \multirow{3}{*}{F: HAVING error}   & Fc1: & Miss HAVING clause                      & {\cellcolor[rgb]{0.984,0.855,0.843}}1   & {\cellcolor[rgb]{0.984,0.855,0.843}}5   & {\cellcolor[rgb]{0.984,0.855,0.843}}12    &{\cellcolor[rgb]{0.984,0.855,0.843}}1    \\
                                   &                                    & Fc2: & Redundant HAVING clause                 & {\cellcolor[rgb]{0.984,0.855,0.843}}0   & {\cellcolor[rgb]{0.984,0.855,0.843}}2   & {\cellcolor[rgb]{0.984,0.855,0.843}}6     &{\cellcolor[rgb]{0.984,0.855,0.843}}1     \\
                                   &                                    & Fe1: & Wrong condition in HAVING               & {\cellcolor[rgb]{0.851,0.851,0.851}}1   & {\cellcolor[rgb]{0.851,0.851,0.851}}2   & {\cellcolor[rgb]{0.851,0.851,0.851}}3     &{\cellcolor[rgb]{0.851,0.851,0.851}}6     \\ 
\hhline{~-------}
                                   & \multirow{2}{*}{G: LIKE error}     & Gc1: & Miss LIKE clause                        & {\cellcolor[rgb]{0.984,0.855,0.843}}1   & {\cellcolor[rgb]{0.984,0.855,0.843}}3   & {\cellcolor[rgb]{0.984,0.855,0.843}}9     &{\cellcolor[rgb]{0.984,0.855,0.843}}4     \\
                                   &                                    & Ge1: & Wrong LIKE condition                    & {\cellcolor[rgb]{0.851,0.851,0.851}}1   & {\cellcolor[rgb]{0.851,0.851,0.851}}8   & {\cellcolor[rgb]{0.851,0.851,0.851}}22    &{\cellcolor[rgb]{0.851,0.851,0.851}}2    \\ 
\hhline{~-------}
                                   & \multirow{2}{*}{H: LIMIT error}    & Hc1: & Redudant LIMIT clause                   & {\cellcolor[rgb]{0.984,0.855,0.843}}1   & {\cellcolor[rgb]{0.984,0.855,0.843}}2   & {\cellcolor[rgb]{0.984,0.855,0.843}}6     &{\cellcolor[rgb]{0.984,0.855,0.843}}0     \\
                                   &                                    & Hc2: & Miss LIMIT clause                       & {\cellcolor[rgb]{0.984,0.855,0.843}}0   & {\cellcolor[rgb]{0.984,0.855,0.843}}1   & {\cellcolor[rgb]{0.984,0.855,0.843}}3     &{\cellcolor[rgb]{0.984,0.855,0.843}}1     \\ 
\hhline{~-------}
                                   & I: INTERSECT error                 & Ie1: & Wrong INTERSECT condition               & {\cellcolor[rgb]{0.851,0.851,0.851}}8   & {\cellcolor[rgb]{0.851,0.851,0.851}}8   & {\cellcolor[rgb]{0.851,0.851,0.851}}9     &{\cellcolor[rgb]{0.851,0.851,0.851}}11     \\ 
\hhline{~-------}
                                   & \multirow{2}{*}{J: DISTINCT error} & Jc1  & Miss a DISTINCT keyword                 & {\cellcolor[rgb]{0.984,0.855,0.843}}7   & {\cellcolor[rgb]{0.984,0.855,0.843}}18  & {\cellcolor[rgb]{0.984,0.855,0.843}}96    &{\cellcolor[rgb]{0.984,0.855,0.843}}12    \\
                                   &                                    & Jc2  & Redundant DISTINCT keyword              & {\cellcolor[rgb]{0.984,0.855,0.843}}4   & {\cellcolor[rgb]{0.984,0.855,0.843}}15  & {\cellcolor[rgb]{0.984,0.855,0.843}}0     &{\cellcolor[rgb]{0.984,0.855,0.843}}11     \\ 
\hhline{~-------}
                                   & K: EXCEPT error                    & Kc1  & Wrong EXCEPT clause                     & {\cellcolor[rgb]{0.984,0.855,0.843}}14  & {\cellcolor[rgb]{0.984,0.855,0.843}}27  & {\cellcolor[rgb]{0.984,0.855,0.843}}24    &{\cellcolor[rgb]{0.984,0.855,0.843}}13    \\ 
\hhline{~-------}
                                   & L: NOT error                       & Lc1  & Miss NOT keyword                        & {\cellcolor[rgb]{0.984,0.855,0.843}}7   & {\cellcolor[rgb]{0.984,0.855,0.843}}9   & {\cellcolor[rgb]{0.984,0.855,0.843}}7     &{\cellcolor[rgb]{0.984,0.855,0.843}}0     \\ 
\hhline{~-------}
                                   & M: UNION                           & Me1  & Wrong UNION condition                   & {\cellcolor[rgb]{0.851,0.851,0.851}}9   & {\cellcolor[rgb]{0.851,0.851,0.851}}9   & {\cellcolor[rgb]{0.851,0.851,0.851}}8     &{\cellcolor[rgb]{0.851,0.851,0.851}10}\\ 
\midrule
\multirow{6}{*}{Semantic errors}   & \multicolumn{3}{l|}{a: Table error}                                                 & {\cellcolor[rgb]{0.98, 0.92, 0.67}}97      & {\cellcolor[rgb]{0.98, 0.92, 0.67}}279          & {\cellcolor[rgb]{0.98, 0.92, 0.67}}274                                       &{\cellcolor[rgb]{0.98, 0.92, 0.67}}53                                       \\
                                   & \multicolumn{3}{l|}{b: Column error}                                                & {\cellcolor[rgb]{0.702,0.808,0.984}}81  & {\cellcolor[rgb]{0.702,0.808,0.984}}237 & {\cellcolor[rgb]{0.702,0.808,0.984}}251   &{\cellcolor[rgb]{0.702,0.808,0.984}}94   \\
                                   & \multicolumn{3}{l|}{c: Miss/redundant Clause/keyword error}                         & {\cellcolor[rgb]{0.984,0.855,0.843}}107 & {\cellcolor[rgb]{0.984,0.855,0.843}}250 & {\cellcolor[rgb]{0.984,0.855,0.843}}432   &{\cellcolor[rgb]{0.984,0.855,0.843}}101   \\
                                   & \multicolumn{3}{l|}{d: Value error}                                                 & {\cellcolor[rgb]{0.82,0.945,0.855}}153  & {\cellcolor[rgb]{0.82,0.945,0.855}}162  & {\cellcolor[rgb]{0.82,0.945,0.855}}230    &{\cellcolor[rgb]{0.82,0.945,0.855}}82    \\
                                   & \multicolumn{3}{l|}{e: Condition error}                                             & {\cellcolor[rgb]{0.851,0.851,0.851}}37  & {\cellcolor[rgb]{0.851,0.851,0.851}}82  & {\cellcolor[rgb]{0.851,0.851,0.851}}88    &{\cellcolor[rgb]{0.851,0.851,0.851}}49    \\
                                   & \multicolumn{3}{l|}{f: Aggregation function error}                                  & {\cellcolor[rgb]{0.737,0.557,0.012}}12  & {\cellcolor[rgb]{0.737,0.557,0.012}}62  & {\cellcolor[rgb]{0.737,0.557,0.012}}25    &{\cellcolor[rgb]{0.737,0.557,0.012}}15    \\
\bottomrule
\end{tabular}}
\caption{The taxonomy of NL2SQL errors with the count of each error type for the models}
\label{tab:taxonomy}
\end{table}

\subsection{NL2SQL error analysis}
\label{sec:error_analysis}
Based on the error taxonomy, we further analyzed the erroneous queries to explore the following three questions:
\begin{itemize}
\item [\textbf{RQ1}] How are the erroneous queries distributed among different models? Do models tend to stumble on the same queries or make mistakes on distinct queries? Furthermore, for those overlapping erroneous queries, do models tend to make similar types of error on them or not? 
\item [\textbf{RQ2}] How do error types spread along the syntactic and semantic dimensions? How different are the distributions of error types among the three models?
\item [\textbf{RQ3}] How far are the erroneous queries from their corresponding ground truths?
\end{itemize}
\subsubsection{The distribution of erroneous queries among models}
\label{sec:error_distribution}
\begin{figure}[!htb]
\centering
\includegraphics[width=0.45\linewidth]{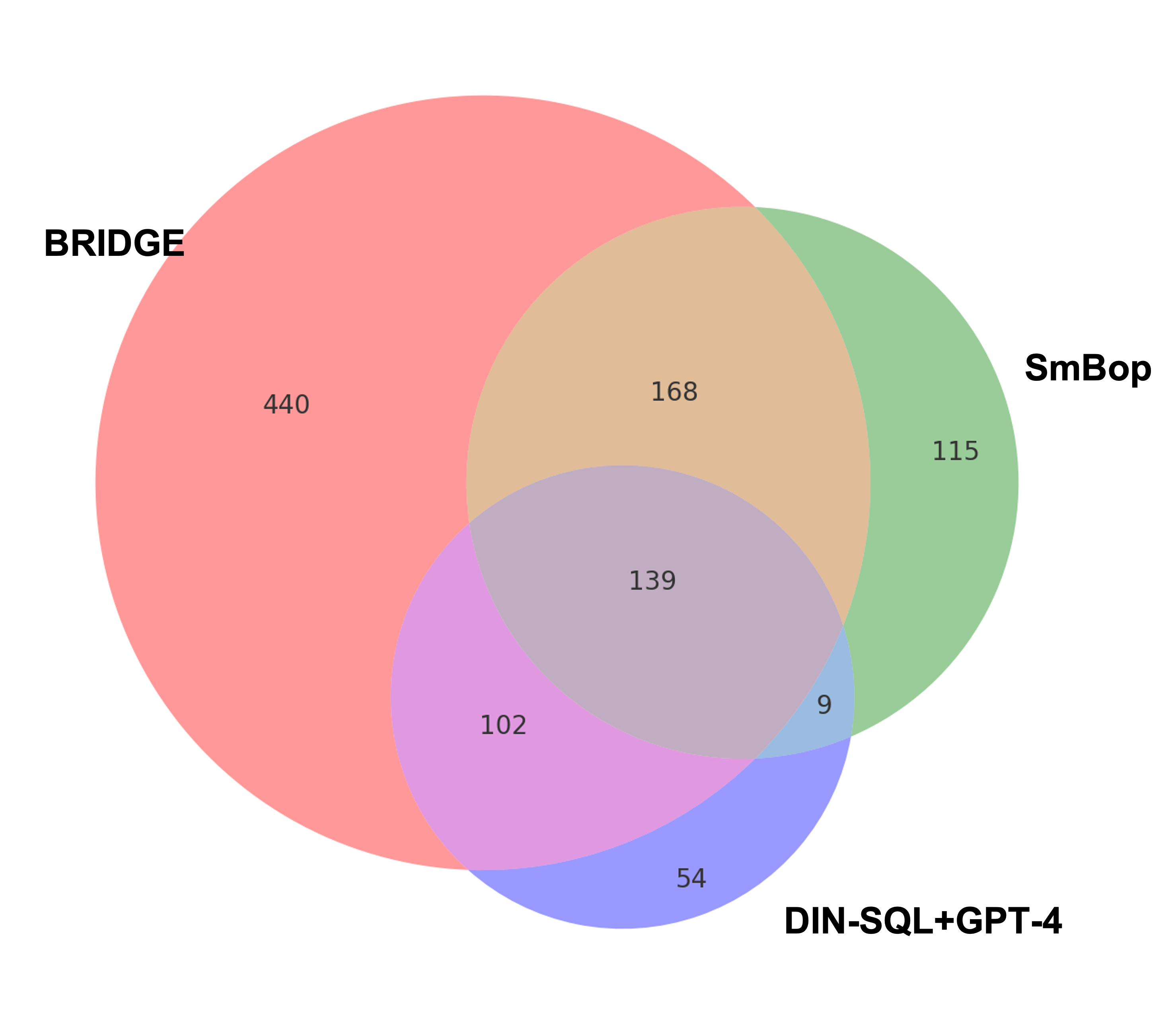}
\caption{The overlap of erroneous queries generated by DIN-SQL+GPT-4, SmBop, and BRIDGE} 
\label{fig:venn}
\end{figure}

Figure~\ref{fig:venn} shows the overlap of erroneous queries among the best three models in a Venn diagram (DIN-SQL+GPT-4, SmBop, and BRIDGE). Each circle represents the queries on which the model made errors. The size of each circle is proportional to the number of erroneous queries of its corresponding model in the sampled dataset (Table~\ref{tab: retrain-stats}). 

Furthermore, we found that there were 129 queries appeared in all four models, we sampled three of them and presented the queries and error types in Table~\ref{tab:commonSQL}. Additionally, we found 92.1\% (280 out of 304) of DIN-SQL+GPT-4's; 82.4\% (355 out of 431) of SmBop's; 84.4\% (720 out of 853) of BRIDGE's; and 70.6\% (750 out of 1062) of GAZP's incorrect queries also confounded other models, indicating that new models are making errors on a limited number of new queries each time. The statistics are also presented in Table~\ref{tab:new_queries}. The results imply that {\bf\em different models tend to make errors on the same subset of queries in NL2SQL.}
\begin{table}[!htb]
    \centering
     \scalebox{0.79}{
    \begin{tabular}{|c|c|} \hline 
         \textbf{Model Names} &  \textbf{Overlapped Queries Percentage}\\ \hline 
         DIN-SQL+GPT-4& 
    92.1\%\\ \hline 
 SmBop&82.4\%\\ \hline 
 BRIDGE&84.4\%\\ \hline 
 GAZP&70.6\%\\ \hline\end{tabular}
 }
    \caption{The percentage of erroneous queries for each model that also appeared in the other three models}
    \label{tab:new_queries}
\end{table}

\begin{equation}
\label{eq:jd}
    Jaccard \; coefficient = \frac{|E_{DIN-SQL+GPT-4} \cap E_{SmBop} \cap E_{BRIDGE} \cap E_{GAZP} |}{| E_{DIN-SQL+GPT-4} \cup E_{SmBop} \cup E_{BRIDGE} \cup E_{GAZP} |}
\end{equation}

To understand whether models make similar types of errors in those overlapped queries, for each query, we measured the similarity of the syntactic and semantic error types respectively among the models. Noticeably, there are multiple distance metrics such as Jaccard distance~\cite{levandowsky1971distance}, Hamming distance, and Euclidean distance. 
\ning{Hamming distance usually works for sets with the same length, while Euclidean distance measures the distance between two points in an Euclidean space. In our case, these two different error sets are challenging to map to the space. Thus, we choose the Jaccard distance as it can be adapted to sets of different lengths and explicitly considers the difference of each error type in the set.}
The Jaccard coefficient is measured using Equation \ref{eq:jd}, where $E_{m}$ means the set of error types that the model $m$ made on the target query. For syntactic error types, 27.9\% of overlapped queries have a Jaccard coefficient of 0 among the four models, which implies that the models did not all make the same syntactic error type in these queries. Regarding the semantic error types, 24.8\% of these queries have a Jaccard coefficient of 0. On the other hand, only 7.8\% of the overlapped queries have the same syntactic error type from the four models, and even fewer of them (6.2\%) have the same semantic error type from all four models. These results show that {\bf\em although the models tend to make errors in the same set of queries, the types of errors in each query tend to be different}. We provide three examples in Table~\ref{tab:commonSQL} to illustrate the similarity and disparity of error types in the same queries.


\begin{table}
\centering
 \scalebox{0.65}{
\begin{tabular}{|c|c|c|c|c|c|c|} 
\hline
& \multicolumn{2}{|c|}{1}                                                                                        & \multicolumn{2}{|c|}{2}                                                                                        & \multicolumn{2}{|c|}{3}                                                                                                                                  \\ 
\hline
\begin{tabular}[c]{@{}c@{}}\textbf{NL query}\end{tabular} & \multicolumn{2}{|c|}{\begin{tabular}[c]{@{}c@{}}How many games are\\ played for all students?\end{tabular}}                                                                  & \multicolumn{2}{|c|}{\begin{tabular}[c]{@{}c@{}}What are the different\\ membership levels?\end{tabular}}                                                                                                                                                                              & \multicolumn{2}{|c|}{\begin{tabular}[c]{@{}c@{}}Find the package choice and series name of the \\TV channel that has high definition TV.\end{tabular}}                      \\ 
\hline
\begin{tabular}[c]{@{}c@{}}\textbf{Correct} \\\textbf{query}\end{tabular}     & \multicolumn{2}{|c|}{\begin{tabular}[c]{@{}c@{}}SELECT sum(gamesplayed) \\FROM Sportsinfo\end{tabular}}       & \multicolumn{2}{|c|}{\begin{tabular}[c]{@{}c@{}}SELECT count(DISTINCT level) \\FROM member\end{tabular}}                                                                & \multicolumn{2}{|c|}{\begin{tabular}[c]{@{}c@{}}SELECT package\_option,  series\_name FROM \\TV\_Channel WHERE high\_definition\_TV = ``yes''\end{tabular}}                   \\ 
\hline
{\cellcolor[rgb]{0.996,0.965,0.965}}\textbf{Model}                                 & \textbf{Generated query}                                                                                 & \textbf{\begin{tabular}[c]{@{}c@{}}Error \\types\end{tabular}} & \textbf{Generated query}                                                                                                                                                                                                                                         & \textbf{\begin{tabular}[c]{@{}c@{}}Error \\types\end{tabular}}                                                  & \textbf{Generated query}                                                                                                                       & \textbf{\begin{tabular}[c]{@{}c@{}}Error \\types\end{tabular}}  \\ 
\hline
{\cellcolor[rgb]{0.996,0.925,0.925}}\begin{tabular}[c]{@{}c@{}}DIN-SQL\\+GPT-4\end{tabular}& \begin{tabular}[c]{@{}c@{}}SELECT COUNT(GameID) \\FROM Plays\_Games~\end{tabular}& \begin{tabular}[c]{@{}c@{}}Da1, Db1, \\Df1\end{tabular}& \begin{tabular}[c]{@{}c@{}}SELECT DISTINCT \\Level FROM member~\end{tabular}& Df2                                                          & \begin{tabular}[c]{@{}c@{}}SELECT package\_option,  series\_name \\FROM TV\_Channel WHERE \\high\_definition\_TV = ``Yes''\end{tabular} & Ad2    \\ \hline 
 {\cellcolor[rgb]{0.996,0.925,0.925}}SmBop                                          & \begin{tabular}[c]{@{}c@{}}SELECT COUNT( * ) , stuid \\FROM plays\_games~\end{tabular}& \begin{tabular}[c]{@{}c@{}}Da1, Db1,\\Db2, Df1\end{tabular}         & \begin{tabular}[c]{@{}c@{}}SELECT Level \\FROM member~\end{tabular}& Df2, Jc1                                                          & \begin{tabular}[c]{@{}c@{}}SELECT package\_option,  series\_name \\FROM TV\_Channel WHERE \\ high\_definition\_TV = 1\end{tabular}                                                   & Ad1    \\ 
\hline
{\cellcolor[rgb]{0.996,0.925,0.925}}BRIDGE                                         & \begin{tabular}[c]{@{}c@{}}SELECT COUNT(*) \\FROM Plays\_Games~\end{tabular}& \begin{tabular}[c]{@{}c@{}}Da1, Db1, \\Df1\end{tabular}         & \begin{tabular}[c]{@{}c@{}}SELECT DISTINCT\\ level FROM member\end{tabular}& Df2                                                     & \begin{tabular}[c]{@{}c@{}}SELECT package\_option,  series\_name \\FROM TV\_Channel WHERE \\ high\_definition\_TV = ``t''\end{tabular}                                                   & Ad4          \\ 
\hline
{\cellcolor[rgb]{0.996,0.925,0.925}}GAZP                                           & \begin{tabular}[c]{@{}c@{}}SELECT count ( * ) \\FROM Plays\_Games\end{tabular}& \begin{tabular}[c]{@{}c@{}}Da1, Db1, \\Df1\end{tabular}         & \begin{tabular}[c]{@{}c@{}}SELECT level \\FROM member~\end{tabular}& \begin{tabular}[c]{@{}c@{}}Df2, Jc1\end{tabular} & \begin{tabular}[c]{@{}c@{}}SELECT package\_option,  series\_name \\FROM TV\_Channel WHERE \\ high\_definition\_TV = ``definition''\end{tabular}                                                   & Ad1          \\ \hline
\end{tabular}}
\caption{Sampled erroneous NL-SQL pairs and their error types}
\label{tab:commonSQL}
\end{table}

\subsubsection{Error frequency}
In this section, we investigate the distribution of error types among models. Specifically, we report the following three measures for each model:
\begin{enumerate}
    \item \textbf{Syntactic error rate ($SYNER_{ms}$)}: Given the model $m$ and a syntactic error type $s$, $SYNER_{ms}$ is the number of queries in which the model $m$ made the syntactic error $s$ divided by the number of ground truth queries in the entire development set that has the corresponding syntax. (Table~\ref{fig:error_percent_rate}). It tells us how likely a syntactical part of a query will produce errors.
    \item \textbf{Syntactic error percentage (distribution) ($SYNEP_{ms}$)}: Given the model $m$ and a syntactic error type $s$, $SYNEP_{ms}$ is the number of queries in which the model $m$ made the syntactic error $s$ divided by the total number of erroneous queries made by the model $m$. (Table~\ref{fig:error_percent_rate}). It measures the percentage of queries that contain a specific type of syntactic error among all erroneous queries.
    \item \textbf{Semantic error percentage (distribution) ($SEMEP_{ms}$)}: Given the model $m$ and the semantic error rate $s$, $SEMEP_{ms}$ is the number of queries in which the model $m$ made the semantic error $s$ divided by the total number of erroneous queries made by the model $m$. (Table~\ref{fig:f_sem}). It measures the percentage of queries that contain a specific type of semantic error among all erroneous queries.
\end{enumerate}


As shown in Table~\ref{fig:error_percent_rate}, the distributions of syntactic error type are similar among all four models. Note that a model can produce a query with multiple types of errors. Notably, the error percentages of \texttt{WHERE}, \texttt{JOIN}, and \texttt{SELECT} are significantly higher than that of other syntactic error types for all the models. However, comparing it with the syntactic error rate, we see that a higher frequency of errors (in all queries) does not equate to a higher error rate when a specific type of keyword is encountered. For example, although \texttt{UNION} errors only account for fewer than 4\% of erroneous queries among all models, it has an error rate of more than 50\% (i.e., when the correct query should contain a \texttt{UNION} clause, the model has a high probability of making errors there). The top 5 syntactic parts that have the highest error rates are shown in Table~\ref{tab:parts}.

\begin{table}[!htb]
\centering
 \scalebox{0.81}{
\begin{tabular}{ccccccccc}
                   & \multicolumn{3}{c}{\textbf{Error Percentage}} &                                                                                                    & \multicolumn{3}{c}{\textbf{Error Rate}} &                                                                                                           \\
Error type         & SmBop                                       & BRIDGE                                      & GAZP                                         &\begin{tabular}[c]{@{}c@{}}DIN-SQL+\\GPT-4\end{tabular}& SmBop                                       & BRIDGE                                      & GAZP                                          &\begin{tabular}[c]{@{}c@{}}DIN-SQL+\\GPT-4\end{tabular}\\ 
\hline
A: WHERE error     & {\cellcolor[rgb]{0.514,0.788,0.878}}47.80\% & {\cellcolor[rgb]{0.627,0.835,0.843}}35.05\% & {\cellcolor[rgb]{0.667,0.847,0.835}}30.89\%  &{\cellcolor[rgb]{0.569,0.808,0.863}}42.43\%& {\cellcolor[rgb]{0.776,0.894,0.804}}18.86\% & {\cellcolor[rgb]{0.698,0.863,0.827}}27.38\% & {\cellcolor[rgb]{0.675,0.851,0.831}}30.04\%   & {\cellcolor[rgb]{0.839,0.918,0.788}}11.81\%\\
B: JOIN error      & {\cellcolor[rgb]{0.784,0.894,0.8}}17.87\%   & {\cellcolor[rgb]{0.69,0.859,0.827}}28.14\%  & {\cellcolor[rgb]{0.659,0.847,0.835}}31.83\%  & {\cellcolor[rgb]{0.765,0.886,0.808}}19.74\%& {\cellcolor[rgb]{0.855,0.925,0.78}}10.13\%  & {\cellcolor[rgb]{0.659,0.847,0.835}}31.58\% & {\cellcolor[rgb]{0.541,0.8,0.867}}44.47\%     & {\cellcolor[rgb]{0.882,0.933,0.776}}7.89\%\\
C: ORDER BY error  & {\cellcolor[rgb]{0.906,0.945,0.769}}4.64\%  & {\cellcolor[rgb]{0.882,0.933,0.776}}7.39\%  & {\cellcolor[rgb]{0.898,0.941,0.773}}5.74\%   & {\cellcolor[rgb]{0.867,0.929,0.78}}9.87\%& {\cellcolor[rgb]{0.906,0.945,0.769}}4.58\%  & {\cellcolor[rgb]{0.816,0.91,0.792}}14.42\%  & {\cellcolor[rgb]{0.82,0.91,0.792}}13.96\%     & {\cellcolor[rgb]{0.878,0.933,0.776}}6.86\%\\
D: SELECT error    & {\cellcolor[rgb]{0.737,0.878,0.816}}23.20\% & {\cellcolor[rgb]{0.714,0.867,0.824}}25.79\% & {\cellcolor[rgb]{0.714,0.867,0.824}}25.80\%  & {\cellcolor[rgb]{0.675,0.851,0.831}}30.92\%& {\cellcolor[rgb]{0.91,0.945,0.769}}4.37\%   & {\cellcolor[rgb]{0.863,0.925,0.78}}9.62\%   & {\cellcolor[rgb]{0.839,0.918,0.788}}11.98\%   & {\cellcolor[rgb]{0.914,0.949,0.765}}4.11\%\\
E: GROUP BY error  & {\cellcolor[rgb]{0.902,0.941,0.769}}5.10\%  & {\cellcolor[rgb]{0.894,0.941,0.773}}5.86\%  & {\cellcolor[rgb]{0.878,0.933,0.776}}7.63\%   & {\cellcolor[rgb]{0.867,0.929,0.78}}9.54\%& {\cellcolor[rgb]{0.91,0.945,0.769}}4.50\%   & {\cellcolor[rgb]{0.855,0.925,0.784}}10.22\% & {\cellcolor[rgb]{0.796,0.902,0.8}}16.56\%     & {\cellcolor[rgb]{0.894,0.941,0.773}}5.93\%\\
F: HAVING error    & {\cellcolor[rgb]{0.945,0.961,0.757}}0.46\%  & {\cellcolor[rgb]{0.941,0.957,0.761}}1.06\%  & {\cellcolor[rgb]{0.929,0.953,0.761}}1.98\%   & {\cellcolor[rgb]{0.925,0.953,0.765}}2.63\%& {\cellcolor[rgb]{0.937,0.957,0.761}}1.44\%  & {\cellcolor[rgb]{0.89,0.937,0.773}}6.47\%   & {\cellcolor[rgb]{0.812,0.906,0.796}}15.11\%   & {\cellcolor[rgb]{0.902,0.941,0.769}}5.76\%\\
G: LIKE error      & {\cellcolor[rgb]{0.945,0.961,0.757}}0.46\%  & {\cellcolor[rgb]{0.937,0.957,0.761}}1.29\%  & {\cellcolor[rgb]{0.922,0.949,0.765}}2.92\%   & {\cellcolor[rgb]{0.929,0.953,0.761}}1.97\%& {\cellcolor[rgb]{0.922,0.949,0.765}}2.86\%  & {\cellcolor[rgb]{0.804,0.902,0.796}}15.71\% & {\cellcolor[rgb]{0.545,0.8,0.867}}44.29\%     & {\cellcolor[rgb]{0.882,0.933,0.776}}8.57\%\\
H: LIMIT error     & {\cellcolor[rgb]{0.945,0.961,0.757}}0.23\%  & {\cellcolor[rgb]{0.945,0.961,0.757}}0.35\%  & {\cellcolor[rgb]{0.941,0.957,0.761}}0.85\%   & {\cellcolor[rgb]{0.945,0.961,0.757}}0.33\%& {\cellcolor[rgb]{0.945,0.961,0.757}}0.41\%  & {\cellcolor[rgb]{0.937,0.957,0.761}}1.24\%  & {\cellcolor[rgb]{0.914,0.949,0.765}}3.73\%    & {\cellcolor[rgb]{0.945,0.961,0.757}}0.41\%\\
I: INTERSECT error & {\cellcolor[rgb]{0.933,0.953,0.761}}1.86\%  & {\cellcolor[rgb]{0.941,0.957,0.761}}0.94\%  & {\cellcolor[rgb]{0.941,0.957,0.761}}0.85\%   & {\cellcolor[rgb]{0.914,0.949,0.765}}3.62\%& {\cellcolor[rgb]{0.78,0.894,0.804}}18.60\%  & {\cellcolor[rgb]{0.78,0.894,0.804}}18.60\%  & {\cellcolor[rgb]{0.757,0.886,0.808}}20.93\%   & {\cellcolor[rgb]{0.714,0.867,0.824}}25.58\%\\
J: DISTINCT error  & {\cellcolor[rgb]{0.925,0.953,0.765}}2.55\%  & {\cellcolor[rgb]{0.914,0.945,0.765}}3.87\%  & {\cellcolor[rgb]{0.867,0.929,0.78}}9.04\%    & {\cellcolor[rgb]{0.882,0.933,0.776}}7.57\%& {\cellcolor[rgb]{0.906,0.945,0.769}}4.78\%  & {\cellcolor[rgb]{0.82,0.91,0.792}}14.35\%   & {\cellcolor[rgb]{0.569,0.808,0.863}}41.74\%   & {\cellcolor[rgb]{0.929,0.953,0.761}}1.74\%\\
K: EXCEPT error    & {\cellcolor[rgb]{0.918,0.949,0.765}}3.25\%  & {\cellcolor[rgb]{0.922,0.949,0.765}}3.17\%  & {\cellcolor[rgb]{0.929,0.953,0.761}}2.26\%   & {\cellcolor[rgb]{0.906,0.945,0.769}}4.28\%& {\cellcolor[rgb]{0.765,0.886,0.808}}20.29\% & {\cellcolor[rgb]{0.592,0.82,0.855}}39.13\%  & {\cellcolor[rgb]{0.631,0.835,0.843}}34.78\%   & {\cellcolor[rgb]{0.765,0.886,0.808}}18.84\%\\
L: NOT error       & {\cellcolor[rgb]{0.933,0.957,0.761}}1.62\%  & {\cellcolor[rgb]{0.941,0.957,0.761}}1.06\%  & {\cellcolor[rgb]{0.941,0.957,0.757}}0.66\%   & {\cellcolor[rgb]{0.945,0.961,0.757}}0.00\%& {\cellcolor[rgb]{0.827,0.914,0.792}}13.46\% & {\cellcolor[rgb]{0.792,0.898,0.8}}17.31\%   & {\cellcolor[rgb]{0.827,0.914,0.792}}13.46\%   & {\cellcolor[rgb]{0.945,0.961,0.757}}0.00\%\\
M: UNION error     & {\cellcolor[rgb]{0.929,0.953,0.761}}2.09\%  & {\cellcolor[rgb]{0.941,0.957,0.761}}1.06\%  & {\cellcolor[rgb]{0.941,0.957,0.757}}0.75\%   & {\cellcolor[rgb]{0.918,0.949,0.765}}3.29\%& {\cellcolor[rgb]{0.435,0.757,0.898}}56.25\% & {\cellcolor[rgb]{0.435,0.757,0.898}}56.25\% & {\cellcolor[rgb]{0.494,0.78,0.882}}50.00\%    &{\cellcolor[rgb]{0.455,0.757,0.998}}62.5\%\\
\hline
\end{tabular}
}
\caption{The error percentage and error rate of each syntactic error type}
\label{fig:error_percent_rate}
\end{table}

Compared to syntactic errors, the distribution of semantic errors is more varied between models (Figure~\ref{fig:f_sem}). We found that \texttt{d: Value error} and \texttt{a: Table error} are the most frequent error types for SmBop (35.5\%) and BRIDGE (32.71\%), respectively. While \texttt{c: Miss/redundant clause/keyword error} appeared most frequently in GAZP (40.68\%) and DIN-SQL+GPT-4 (33.22\%). It indicates that {\bf\em the semantic challenges of the investigated NL2SQL models are more varied than the syntactic challenges they faced.}

\begin{table}[!htb]
\centering
 \scalebox{0.81}{
\begin{tabular}{ccccc}
\textbf{Error type}                              & \textbf{SmBop}                                       & \textbf{BRIDGE}                                      & \textbf{GAZP}                                          &\textbf{\begin{tabular}[c]{@{}c@{}}DIN-SQL\\+GPT-4\end{tabular}}\\ 
\hline
a: Table error                          & {\cellcolor[rgb]{0.667,0.847,0.835}}22.51\% & {\cellcolor[rgb]{0.537,0.796,0.871}}32.71\% & {\cellcolor[rgb]{0.624,0.831,0.847}}25.80\%   & {\cellcolor[rgb]{0.784,0.894,0.8}}17.43\%\\
b: Column error                         & {\cellcolor[rgb]{0.71,0.867,0.824}}18.79\%  & {\cellcolor[rgb]{0.6,0.82,0.855}}27.78\%    & {\cellcolor[rgb]{0.651,0.843,0.839}}23.63\%   & {\cellcolor[rgb]{0.675,0.851,0.831}}30.92\%\\
c: Miss/redundant Clause/keyword error~ & {\cellcolor[rgb]{0.635,0.835,0.843}}24.83\% & {\cellcolor[rgb]{0.58,0.816,0.859}}29.31\%  & {\cellcolor[rgb]{0.435,0.757,0.898}}40.68\%   & {\cellcolor[rgb]{0.631,0.835,0.843}}33.22\%\\
d: Value error                          & {\cellcolor[rgb]{0.502,0.784,0.878}}35.50\% & {\cellcolor[rgb]{0.71,0.867,0.824}}18.99\%  & {\cellcolor[rgb]{0.675,0.851,0.831}}21.66\%   & {\cellcolor[rgb]{0.69,0.859,0.827}}26.97\%\\
e: Condition error                      & {\cellcolor[rgb]{0.839,0.918,0.788}}8.58\%  & {\cellcolor[rgb]{0.827,0.914,0.792}}9.61\%  & {\cellcolor[rgb]{0.843,0.918,0.784}}8.29\%    & {\cellcolor[rgb]{0.796,0.902,0.8}}16.12\%\\
f: Aggregation function error           & {\cellcolor[rgb]{0.914,0.945,0.765}}2.78\%  & {\cellcolor[rgb]{0.859,0.925,0.78}}7.27\%   & {\cellcolor[rgb]{0.91,0.933,0.808}}2.35\%    & {\cellcolor[rgb]{0.906,0.945,0.769}}4.93\%\\
\hline
\end{tabular}
}
\caption{The error percentage of each semantic error type}
\label{fig:f_sem}
\end{table}


\begin{table*}
\centering
\scalebox{0.75}{
\begin{tabular}{cccccc} 
\toprule
\textbf{Model}  & \textbf{Top 1} & \textbf{Top 2} & \textbf{Top 3} & \textbf{Top 4} & \textbf{Top 5}  \\ 
\hline
\textbf{SmBop}  & UNION 56.25\%             & EXCEPT 20.29\%            & WHERE 18.86\%             & INTERSECT 18.60\%         & NOT 13.46\%                \\ 
\textbf{BRIDGE} & UNION 56.25\%             & EXCEPT 39.13\%            & JOIN 31.58\%              & WHERE 27.38\%             & INTERSECT 18.6\%           \\ 
\textbf{GAZP}   & UNION 50.00\%             & JOIN 44.47\%              & LIKE 44.29\%              & DISTINCT 41.74\%          & EXCEPT 34.78\%             \\
\textbf{\begin{tabular}[c]{@{}c@{}}DIN-SQL+GPT-4\end{tabular}} & UNION 62.5\%    & INTERSECT 25.58\%  & EXCEPT 18.84\%    &WHERE 11.81\% &    LIKE 8.57\%\\
\bottomrule
\end{tabular}
}
\caption{The top 5 error-prone syntactic parts of a SQL query for the selected models}
\label{tab:parts}
\end{table*}


\begin{figure}[!h]
\centering
\includegraphics[width=0.9\linewidth]{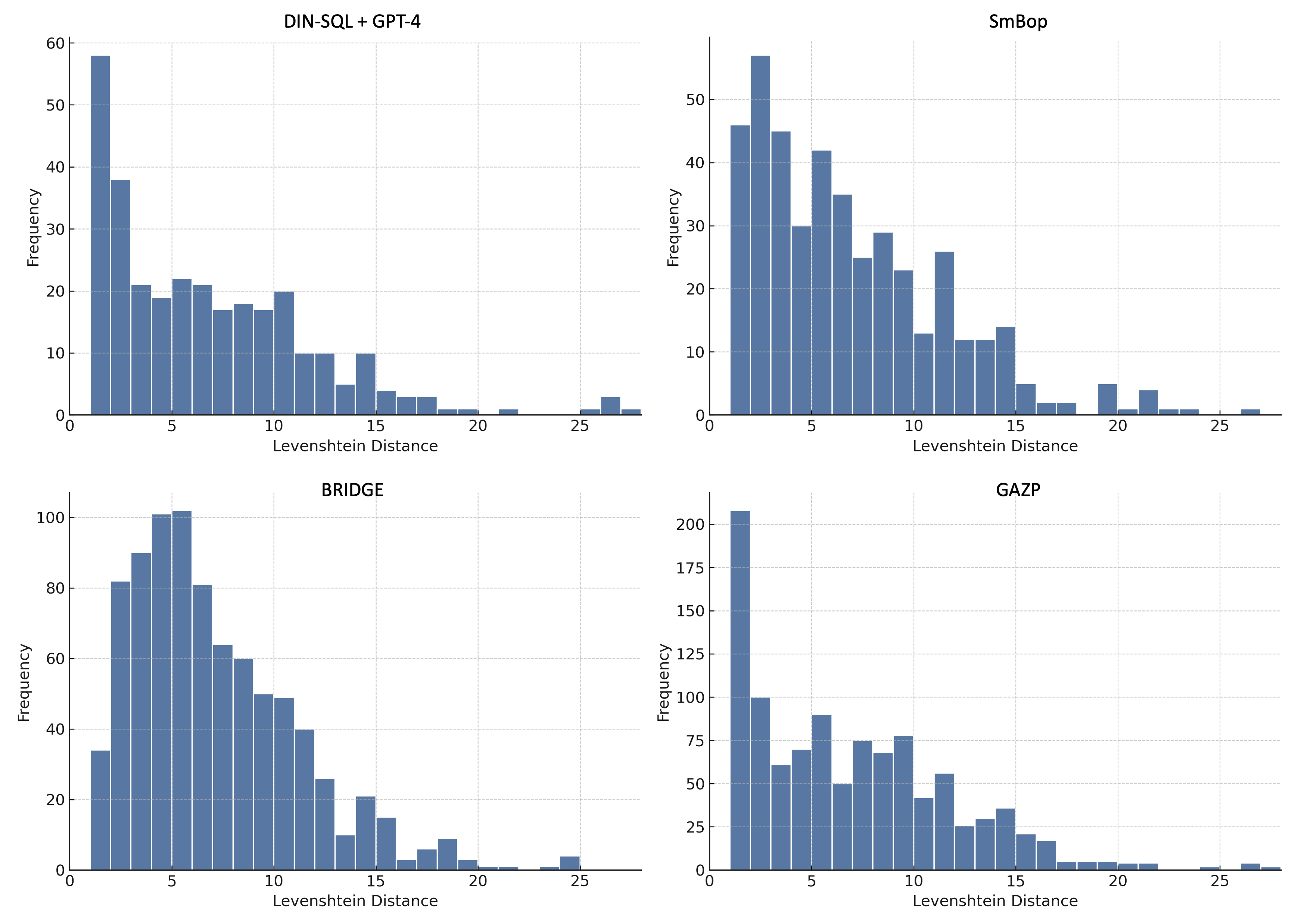}
\caption{The distribution of Levenshtein distances between erroneous queries and ground truth queries for each model}
\label{fig:lev}
\end{figure}

\subsubsection{Distance between erroneous and ground truth queries}
Lastly, we used Levenshtein distance to measure the distance between erroneous and ground truth queries. The Levenshtein distance between two queries is defined as the minimum number of word-level (split by space) edits (insertions, deletions, or substitutions) required to transform the model-generated query into the ground truth query. Before computing the distance, we first pre-process the ground truth and predicted queries for each pair to unify the SQL format. Specifically, we i): ignore the differences in the upper or lower case letters except for values; ii) extract the table names and the alias for each column used in the query and prefix the column names with the identified table names. Noticeably, there may be other cases where the predicted query does not need to be revised to exactly the form of the ground truth query in order to get the correct result; therefore, the Levenshtein distances we obtained may be larger than the actual ones.

Figure~\ref{fig:lev} shows the distribution of the Levenshtein distances of errors made by each model. It is worth noting that all four distributions have a long tail. Specifically, by looking at the Levenshtein distance in three different groups: 0--5, 6---10, and more than 10; we found that a large portion of erroneous SQL queries for all four models can be fixed in a small number of edits. Especially for the best model we reproduced, DIN-SQL+GPT-4, 19.1\% (58/304) of the erroneous queries can be fixed in one step. In particular, 19.6\% (208/1062) queries in GAZP only need changes in one token; the percentage of erroneous queries requiring only changes in one token for BRIDGE and SmBop is 3.9\% (34/853) and 10.7\% (46/431), respectively.\looseness=-1


\section{Analysis on NL2SQL Human-Model Attention Alignment}
After understanding the error types and distribution in NL2SQL, we aim to further investigate the possible cause of SQL errors.
Previous research~\cite{attention1, attention6, attention3} has shown that the discrepancy between model attention and human attention is correlated with poor performance in code summarization, machine translation, and visual feature representation.
We hypothesize that such misalignment between model attention and human attention is also a source of error in NL2SQL. In other words, NL2SQL models return incorrect columns or tables in part because they do not pay attention to words that humans pay attention to. However, the lack of empirical research limits our understanding in this area. \looseness=-1


To investigate whether and to what extent SQL errors derive from such attention misalignment, we compare human-labeled attention with a representative model's attention on NL queries. 

\subsection{Data preparation}

\subsubsection{Different attention calculation methods}
There are different methods to obtain the model's attention. For example, some work~\cite{clark-etal-2019-bert, Galassi_2021, DBLP:journals/corr/LiMJ16a, DBLP:journals/corr/abs-1909-11218} leveraged the self-attention mechanism of transformer~\cite{attention_is_all_you_need} to obtain attention, while some work~\cite{DBLP:journals/corr/DenilDF14, 10.5555/3305890.3306006, simonyan2014deep, sundararajan2017axiomatic} used the gradients of the model predictions with respect to the input features to calculate the model’s attention. 

Perturbation-based methods follow a two-step process: they first mutate the input and then calculate the model's attention based on the differences in the output. For instance, LIME~\cite{ribeiro-etal-2016-trust} generates a local explanation by approximating the specific model predictions using a simpler model, such as a linear classifier. SHAP~\cite{10.5555/3295222.3295230} improves on LIME by perturbing the input based on game theory and using the Shapely value to estimate the importance of different tokens. However, a limitation of both methods is that they often require a large number of perturbed samples to ensure an accurate estimation.
Furthermore, LIME and SHAP only mutate an input by deleting tokens, which may significantly alter the meaning or structure of the input. To address this limitation, more recent perturbation-based methods opt to replace tokens with similar or semantically related tokens in the context~\cite{10.1109/TVCG.2018.2865230, wu-etal-2020-perturbed}. These methods often utilize a masked language model such as BERT~\cite{devlin-etal-2019-bert} to predict similar or semantically related tokens to replace existing tokens in an input. 
In this study, we selected a perturbation-based method optimized by BERT. The reasons for this choice will be explained in the following paragraphs.

\subsubsection{NL2SQL model selection}
We selected SmBoP~\cite{rubin2020smbop}, which achieves the second-highest performance among the models we used in the previous study. Although DIN-SQL + GPT-4~\cite{pourreza2023din} achieves the highest performance, GPT-4 is not open-source, therefore, we cannot directly calculate its attention based on self-attention or gradient. Additionally, by using the perturbation-based method, the GPT-4 API has to be called millions of times, which is not practical and affordable.

\subsubsection{Attention calculation method selection}
Methods leveraging either self-attention or gradient are not suitable for a model with multiple components other than a transformer~\cite{attention_is_all_you_need} since attention or gradient does not represent the attention for the entire model.
The decoding process of SmBoP starts by generating SQL components and gradually combines them from the bottom up to create a complete SQL statement. This process involves contextualization based on the transformer's self-attention mechanism and the use of predefined rules to filter out invalid queries.

Due to the complexity of SmBoP architecture, the attention in transformer headers or simply using gradients cannot accurately represent the attention of the entire model. 
Therefore, we choose the perturbation-based method, which regards the SmBoP model as a black box without considering the inner details of its architecture.

\subsection{Human attention labeling}
To obtain human attention on NL2SQL tasks, two annotators, who are proficient in SQL, conducted iterative refinements of attention annotation.
We first randomly sampled 200 tasks from the Spider dataset, where 140 tasks can be correctly solved by the experiment model (SmBoP), and 60 tasks on which the model makes errors. We intentionally balanced the sampled tasks according to the performance of SmBoP (69.5\% on the Spider test set).
 
Each iteration consisted of the following three steps. First, two annotators separately annotated a new sample batch of 25 tasks. For each task, the annotators reviewed the natural language (NL) question and the ground truth SQL query. Then each annotator individually identified important NL words (those that contribute to the query) and marked their attention weight as \textbf{1} while marking the attention weight of the remaining unimportant words (e.g., all, the, of) as \textbf{0}.
Second, we computed the inter-rater reliability between annotators~\cite{10.1145/3359174} (Fleiss' Kappa and Krippendorff's Alpha) at the end of each iteration.
Lastly, annotators discussed how they judge the importance of a certain word.
Annotators completed three refinement iterations until the human-labeled attention became stable and the inter-rater reliability scores were substantial. At the end of the final refinement iteration, Fleiss' Kappa was 0.67 and Krippendorff's Alpha was 0.58.

\subsection{Measuring attention alignment between human and model}
We measure the alignment between the attention of the models and the annotators using the \textit{ keyword coverage rate}.
Specifically, we select the Top $K$ model-focused words with the highest attention as $word_m$, where $K$ equals the number of keywords selected by human annotators.
Then we calculate the percentage of human-labeled words $word_h$ covered in $word_m$, as shown in Equation~\ref{eq:keyword_coverage_rate}.

\begin{equation}
    rate = \frac{|word_m \cap word_h|}{|word_h|}
    \label{eq:keyword_coverage_rate}
\end{equation}

\vspace{3mm}
Nevertheless, the attention scores calculated by this method cannot be directly compared to words annotated by human labelers. This is because human labelers annotate attention on each individual word, while machine attention is calculated and distributed on specific tokenization of the model. For example, the word ``apple'' can be tokenized into two tokens: ``ap'' and ``ple'' through byte pair encoding.
To bridge the gap, we developed a method to map model tokens back to individual words and recalculate the attention distribution. 
Suppose an input text includes $N$ words $\{w_1, w_2,...,w_n\}$, while the model's tokenizer splits the text into $M$ tokens $\{t_1, t_2,...,t_m\}$. We calculate the model’s attention on $i$th word $w_i$ as the sum of all tokens that overlap with $w_i$, as shown in Equation~\ref{eq:token_map}.

\begin{equation}
    attention_{w_i} = \sum^{}_{ t_j \cap w_i \neq \varnothing } attention_{t_j}
    \label{eq:token_map}
\end{equation}

\subsection{Results}

In this section, we aim to answer the following research questions: 
\begin{itemize}
    \item[RQ1] To what extent model attention is aligned with human attention?
    \item[RQ2] Can the attention misalignment explain the errorous queries generated by NL2SQL models?
    \item[RQ3] Which error types are highly related to the attention misalignment?
\end{itemize}

In our 200 sampled tasks from the Spider dataset, the average number of words in the NL queries is $12$.
Accordingly, we calculate the \textit{Keyword Coverage Rate} by experimenting with different $K$ (i.e., the top $K$ words with the highest attention to be considered in the calculation) from 1 to 20.
Since the value of $K$ may exceed the number of words in the NL query, the actual number of words considered is equal to the minimum of $K$ and the number of words in the NL query.

\begin{table}[!h]
\centering
\begin{tabular}{c|c|c|c|c|c|c|c|c|c|c}
\hline
\textbf{\textit{K}} & \textbf{1} & \textbf{2} & \textbf{3} & \textbf{4} & \textbf{5} & \textbf{6} & \textbf{7} & \textbf{8} & \textbf{9} & \textbf{10} \\
\hline
\textbf{\textit{Alignment}} & 0.138 & 0.231 & 0.302 & 0.368 & 0.433 & 0.491 & 0.545 & 0.601 & 0.649 & 0.698 \\
\hline
\hline
\textbf{\textit{K}} & \textbf{11} & \textbf{12} & \textbf{13} & \textbf{14} & \textbf{15} & \textbf{16} & \textbf{17} & \textbf{18} & \textbf{19} & \textbf{20} \\
\hline
\textbf{\textit{Alignment}} & 0.740 & 0.776 & 0.804 & 0.832 & 0.851 & 0.868 & 0.880 & 0.888 & 0.894 & 0.898 \\
\hline
\end{tabular}
\caption{Average Keyword Coverage Rate for all 200 tasks}
\label{tab:distribution_alignment}
\end{table}

\begin{figure*}[!h]
    \centering
    \includegraphics[width=0.9\linewidth]{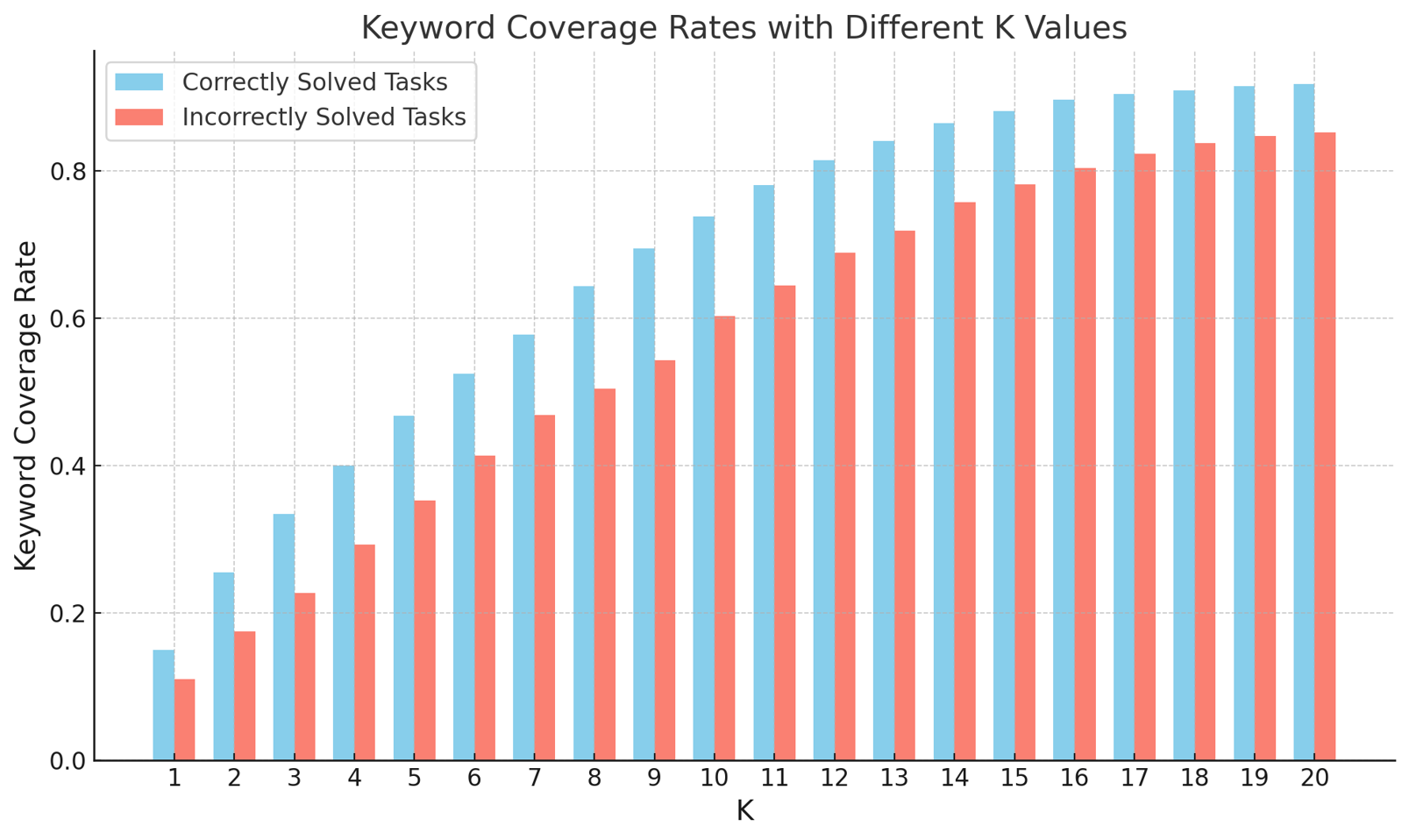}
    \caption{The distribution of alignment for correctly and incorrectly solved tasks considering different numbers of keywords.}
    \label{fig:different_K}
\end{figure*}

\paragraph{\textbf{F1: The model's attention partially aligns with human attention, which is consistent with the model's performance.}}
Table~\ref{tab:distribution_alignment} shows the average \textit{Keyword Coverage Rate} for all 200 tasks in different settings of $K$. For example, when considering the top 12 words (the average number of words), there is an overlap of around 77.6\% between human-focused and model-focused words. This result is consistent with the performance of SmBoP, which achieves 70\% accuracy on the sampled dataset.

\paragraph{\textbf{F2: Attention alignment is higher when the model correctly solves the task, suggesting that SQL errors are correlated with attention misalignment.}}
To examine the relationship between attention alignment and the model's performance, we compare the \textit{Keyword Coverage Rate} of correctly solved tasks with that of incorrectly solved tasks.
Figure~\ref{fig:different_K} shows the average keyword coverage rate with different $K$ values for correct and incorrect queries generated by SmBoP. When SmBoP generates a correct SQL query, the attention alignment is significantly higher than when it generates an erroneous SQL for all $K$ values, with all $p$-values being less than $0.05$.

\begin{figure*}[!h]
    \centering
    \includegraphics[width=0.8\linewidth]{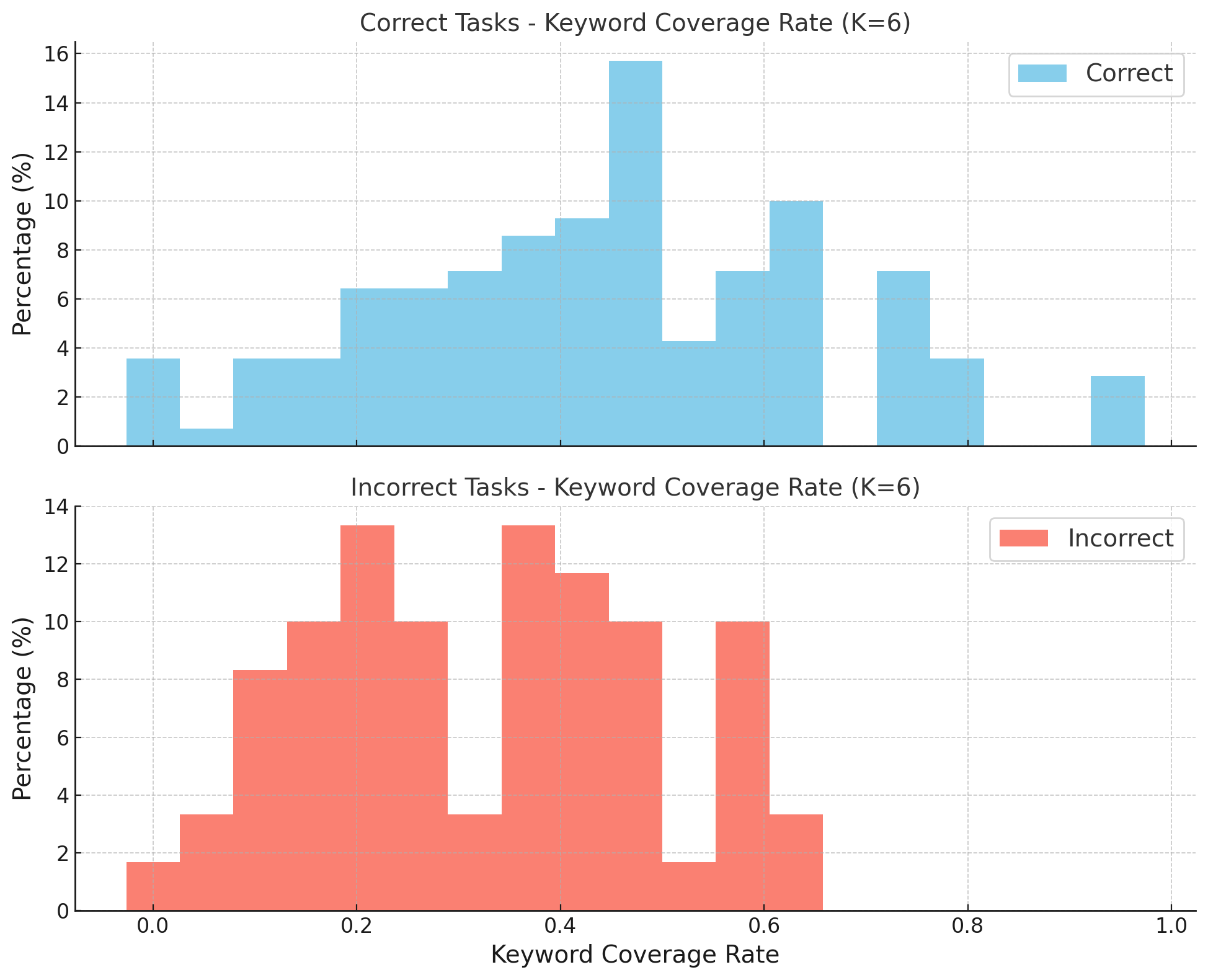}
    \caption{A comparison of attention alignment between correctly and incorrectly solved tasks.}
    \label{fig:attention_alignment}
\end{figure*}

Take $K=6$ (half of the number of words) as an example, Figure~\ref{fig:attention_alignment} shows the comparison of attention alignment distributions between correctly and incorrectly generated queries. Each distribution approximately follows a Gaussian distribution. However, the mean of the distribution for erroneous queries is significantly lower than that of correct queries, with a $p$-value of $1.5e-4$. Furthermore, when generated queries are correct, all the rates are no more than $0.67$, suggesting that a low attention alignment contributes to generating an erroneous SQL.

\paragraph{\textbf{F3: Attention misalignment is not specific to a certain error type.}}
To further investigate which types of errors are more correlated with attention misalignment, we labeled the error types of all 40 incorrectly solved tasks using the same procedure discussed in Section~\ref{sec:coding}. Next, we calculated the average keyword coverage rate associated with each type of error. Specifically, for a certain error type, we summed up the attention alignment of all tasks that included that error. Then, we divided the summed attention alignment score by the number of tasks that included this error, as shown in Equation~\ref{eq:keyword_coverage_rate}.

\begin{equation}
    align_{E_i} = \frac{\sum^{}_{ E_i \in task_j} align_{task_j}}{\left| \{ \text{{task}}_i \mid \text{{task}}_j \text{{ contains error }} E_i \} \right|}
    \label{eq:token_map}
\end{equation}

Table~\ref{tab:attention_error} shows the average alignment for each type of error. The results indicate no significant difference in attention alignment across different types of error.

\begin{table}[h]
\centering
\caption{Average attention alignment for different types of error.}
\label{tab:attention_error}
\[
\begin{array}{c|c|c|c|c|c|c|c}
\hline
\textbf{Syntactic error type} & A & B & C & D & E & F & G \\
\hline
\text{Attention alignment} & 0.26 & 0.32 & 0.22 & 0.30 & 0.25 & 0.27 & 0.34 \\
\hline
\textbf{Syntactic error type} & H & I & J & K & L & M \\
\hline
\text{Attention alignment} & 0.19 & 0.28 & 0.33 & 0.24 & 0.21 & 0.26 \\
\hline
\end{array}
\]
\end{table}

\section{The User Study of Interactive Error Discovery \& Repair Mechanisms}
\label{sec:study_error_handling_mechanisms}
In the past few years, we have seen a growing interest in interactive mechanisms for users to detect and repair NL2SQL errors~\cite{narechania2021diy, SQLVis, misp, wang2021interactive, gur2018dialsql, leventidis2020queryvis, bergamaschi2013quest}, regardless of users' domain expertise of using SQL language. To understand the performance and usage of these mechanisms by users, we conducted a controlled user study to examine the effectiveness of different error discovery and repair mechanisms for NL2SQL\footnote{The protocol of the study has been reviewed and approved by the IRB at our institution.}. Specifically, we investigated the following research questions.
\begin{enumerate}
    \item[\textbf{RQ1.}] How effective and efficient are the different error-handling mechanisms and interaction strategies in NL2SQL? 
    \item[\textbf{RQ2.}] What are the user preferences and perceptions of different mechanisms and strategies?
    \item[\textbf{RQ3.}] What are the gaps between the capabilities of existing approaches and user needs?
\end{enumerate}



\subsection{Experiment conditions}
\label{sec:strategies}

In this study, we used four conditions shown in Table~\ref{tab:conditions}. In the baseline condition, no interactive support was provided for error discovery and repair. Users had to examine the correctness of a generated SQL query by directly checking the query result and manually editing a generated SQL query to fix an error.

In addition to the baseline, we selected three experimental conditions based on three representative approaches for error discovery and repair. The first experimental condition exemplifies an \textbf{explanation- and example-based} approach (DIY~\cite{narechania2021diy}) that displays intermediate results by decomposing a long SQL query into shorter queries and generating natural language explanations for each step. Meanwhile, it allows users to fix the mapping between words in the NL description and their corresponding entities in the generated SQL query from a drop-down menu. The second experimental condition uses an \textbf{explanation-based visualization} approach (SQLVis~\cite{SQLVis}). The technique uses a graph-based visualization for the generated SQL query to illustrate the explicit and implicit relationship among different SQL components such as the selected columns, tables, primary and foreign keys. The third experimental condition exemplifies a \textbf{conversational dialog approach} (MISP~\cite{misp}). It allows users to correct an erroneous SQL query through multiple rounds of conversation in natural language (Table~\ref{tab:conditions}).
We replicated the core functionalities of the DIY mechanism used in experimental condition \#1 as described in the paper~\cite{narechania2021diy} because the official source code was not publicly released. For experimental condition \#2, we used the official implementation\footnote{\url{https://github.com/Giraphne/sqlvis}} provided by the authors. For the dialog system under experimental condition \#3, we implemented an interactive widget based on the open-sourced command-line tool and an interactive graphical user interface based on the React-Chatbot-Kit\footnote{\url{https://www.npmjs.com/package/react-chatbot-kit}} for the study. \looseness=-1

\begin{table}
\centering
 \scalebox{0.9}{
\begin{tabular}{c|c} 
\toprule
\textbf{Condition} & \textbf{Error Discovery and Repair Mechanisms}  \\ 
\hline
Baseline           & Direct SQL query editing      \\ 
\hline
Exp. Cond. \#1    & Step-by-step SQL query explanation \& NL-SQL entity mapping (DIY~\cite{narechania2021diy})        \\ 
\hline
Exp. Cond. \#2          & Graph-based SQL query visualization (SQLViz~\cite{SQLVis})                       \\
\hline
Exp. Cond. \#3          & Conversational dialog system (MISP~\cite{misp})   \\ 
\bottomrule
\end{tabular}
}
\caption{The list of conditions used in the user study}
\label{tab:conditions}
\end{table}

\subsection{Participants}
We recruited 26 participants from the campus community of a private university in the midwest of the United States through mailing lists and social media. Participants included 15 men and 11 women aged 20 to 30 years. Nine participants were novice SQL users who had either no experience in using SQL or had seen SQL queries before but were not familiar with the syntax. 10 participants were intermediate SQL users who had either taken an introductory database course or understood the SQL syntax. The remaining 7 were experienced users who were familiar with SQL queries or had significant experience working with SQL. Each participant was compensated with \$15 USD for their time.\looseness=-1


\subsection{Study procedure}
In our study, each participant experienced the four conditions described in Section~\ref{sec:strategies}. As the goal of this study is to investigate the error discovery and repair behavior of users, the example SQL queries for each participant were randomly selected from the dataset of incorrect queries generated by the three NL2SQL models used in the error analysis study. Each query that a participant encountered was also randomly assigned to one of the experimental conditions or the baseline condition. \looseness=-1

To facilitate the user experiment, we implemented a web application that can automatically select SQL tasks and assign conditions to study participants. After finishing one SQL query, users can click the ``Next'' button on the application, and it will randomly select the next query and assign a condition to it. Both the query assignment and the condition assignment were randomized. For each query, the web application renders the task description, the database and its tables, and the assigned error-handling mechanisms.\looseness=-1

Each experiment session began with the informed consent process. Then, each participant watched a tutorial video about how to interact with the system to solve an SQL task and fix NL2SQL errors under different conditions. Then, each participant was given a total of 45 minutes to solve as many SQL tasks as possible. On average, each participant completed 22.0 SQL tasks in 45 minutes (5.5 in each condition). After each experiment session, the participant completed a post-study questionnaire. This questionnaire asked participants to rate their overall experience, the usefulness of interactive tool support under different conditions, and their preferences in Likert scale questions. We ended each experiment session with a 10-minute semi-structured interview. In the interview, we asked follow-up questions about their responses to the post-study questionnaire, if they encountered any difficulties with interaction mechanisms under the conditions, and which parts they found useful. We also asked participants about the general workflow as they approached the task and the features they wished they had when handling NL2SQL errors. All user study sessions were video recorded with the consent of the participants.

Following established open coding methods~\cite{braun2006using,lazar2017research}, an author conducted a thematic analysis of the interview transcripts to identify common themes about user experiences and challenges they encountered while using the different error handling mechanisms, as well as their suggestions for new features. Specifically, the coder went through and coded the transcripts of the interview sessions using an inductive approach. For user quotes that did not include straightforward key terms, the coder assigned researcher-denoted concepts as the code.

\subsection{Data collection}
For each SQL task, we collected three types of data from the participant: (1) the updated SQL query after their repair; (2) the starting and ending time; (3) the user's interaction log with the error handling mechanism (e.g., clicking to view the sampled table, opening the drop-down menu, interacting with the chatbot).

We cleaned up the data from the participants through the following steps. First, we filtered out the queries that are skipped by the participants (i.e., the user clicking on ``Next'' without making any changes to the query), which consist of less than 10\% of the total data. Second, if the participant did not utilize the interaction mechanism associated with the experimental condition at all (e.g., the user inspected the query without using any assistance and modified the query manually), the task was deemed to be solved using the baseline method.

\subsection{Results}
\label{sec:findings}
In this section, we report the key findings on the efficiency, effectiveness, and usability of different error handling mechanisms and their user experiences. For each condition in a statistical test, the data is sampled evenly and randomly. \looseness=-1

\paragraph{\textbf{F1: The error handling mechanisms do not significantly improve the accuracy of fixing erroneous SQL queries}} To start with, we conducted a one-way ANOVA test ($\alpha$=0.05) among tasks that used different error handling mechanisms. The p-value for the accuracy was 0.82, indicating that there were no significant differences between the different error handling methods. The average accuracy and standard deviation among the participants are shown in Table~\ref{tab:cond-acc-toc}. \looseness=-1

\begin{table}
\centering
\setlength{\extrarowheight}{0pt}
\addtolength{\extrarowheight}{\aboverulesep}
\addtolength{\extrarowheight}{\belowrulesep}
\setlength{\aboverulesep}{0pt}
\setlength{\belowrulesep}{0pt}
\resizebox{0.75\linewidth}{!}{
\begin{tabular}{c|cc|cc} 
\toprule
\textbf{Conditions 
} & \textbf{Avg. Acc. ($\mu=0.56$)}                       & \textbf{SD ($\mu=0.50$)}                    & \textbf{Avg. ToC ($\mu=116.7$)}                      & \textbf{SD ($\mu=89.6$)}                             \\ 
\hline
\textbf{B1}         &{\cellcolor[rgb]{0.96862745, 0.8980392, 0.8980392}}0.55                                     & {\cellcolor[rgb]{0.8,0.9843137,0.5647059}}0.48                                    & {\cellcolor[rgb]{0.78431374,0.9882353,0.53333336}}109.7                                & {\cellcolor[rgb]{0.93333334,0.7607843,0.75686276}}95.8                 \\
\textbf{C1}         & {\cellcolor[rgb]{0.9843137, 0.95686275, 0.9607843}}0.56 & 
{\cellcolor[rgb]{0.9647059,0.8745098,0.8745098}}0.51                                    &{\cellcolor[rgb]{0.8117647,0.99215686,0.59607846}}110.9  & {\cellcolor[rgb]{0.9098039,0.68235296,0.6745098}}101.0  \\
\textbf{C2}         & {\cellcolor[rgb]{0.80784315, 0.9882353, 0.5882353}}0.60 & 
{\cellcolor[rgb]{0.99607843,0.99607843,0.9882353}}0.50                                    & {\cellcolor[rgb]{0.92941177,0.99607843,0.85490197}}115.9 & {\cellcolor[rgb]{0.92941177,0.7529412,0.7529412}}96.5  \\
\textbf{C3}         & {\cellcolor[rgb]{0.93333334, 0.7647059, 0.7607843}}0.53     & 
{\cellcolor[rgb]{0.9647059,0.8745098,0.8745098}}0.51 & {\cellcolor[rgb]{0.9254902,0.73333335,0.7294118}}128.5 & {\cellcolor[rgb]{0.7490196,0.9843137,0.45490196}}61.7  \\

\bottomrule
\end{tabular}}
\caption{The average accuracy and ToC (in seconds) for different conditions}
\label{tab:cond-acc-toc}
\end{table}

We then analyzed the effect of different mechanisms on the accuracy of fixing specific error types, including five common syntactic error types (\texttt{A: WHERE error; B: JOIN error; C: ORDER BY error; D: SELECT error; E: GROUP BY error}, as well as six semantic errors shown in Table~\ref{fig:error_percent_rate}. Using the same statistical test, we found that the p-values for all types of error were higher than the 0.05 threshold, indicating that there were no significant differences in accuracy when the user used different error-handling mechanisms (Table~\ref{tab:cond-errorType-accuracy}).

\begin{table}[!htb]
\centering
\resizebox{0.7\linewidth}{!}{
\begin{tabular}{c|ccccc|cccccc} 
\toprule
                   & \multicolumn{5}{c|}{\textbf{Syntactic types 
                   }}                  & \multicolumn{6}{c}{\textbf{Semantic types 
                   }}                                  \\
                   & \textbf{A} & \textbf{B} & \textbf{C} & \textbf{D} & \textbf{E} & \textbf{a} & \textbf{b} & \textbf{c} & \textbf{d} & \textbf{e} & \textbf{f}  \\ 
\hline
\textbf{B1}        &{\cellcolor[rgb]{0.9254902, 0.7411765, 0.7411765}}0.42            &{\cellcolor[rgb]{0.9411765, 0.99607843,0.88235295}}0.40           &{\cellcolor[rgb]{0.78039217, 0.9882353, 0.52156866}}0.38            &{\cellcolor[rgb]{0.78039217, 0.9882353, 0.52156866}}0.67            &{\cellcolor[rgb]{0.9764706, 0.9137255, 0.9098039}}0.29           &{\cellcolor[rgb]{0.84705883, 0.99215686, 0.6784314}}0.40           &{\cellcolor[rgb]{0.8352941, 0.99607843, 0.6392157}}0.58           &{\cellcolor[rgb]{0.81960785, 0.9882353, 0.6039216}}0.52           &{\cellcolor[rgb]{0.7529412,0.9843137, 0.4745098}}0.45           &{\cellcolor[rgb]{0.96862745, 0.8980392, 0.8980392}}0.32            &{\cellcolor[rgb]{0.972549,0.89411765,0.8862745}}0.25            \\
\textbf{C1}        &{\cellcolor[rgb]{0.91764706, 0.6901961, 0.68235296}}0.40            &{\cellcolor[rgb]{0.78039217, 0.9882353, 0.52156866}}0.44           &{\cellcolor[rgb]{0.92941177, 0.73333335, 0.7254902}}0.27           &{\cellcolor[rgb]{0.92941177, 0.73333335, 0.7254902}}0.56           &{\cellcolor[rgb]{0.9254902, 0.7176471, 0.70980394}}0.24           &{\cellcolor[rgb]{0.77254903, 0.9882353, 0.50980395}}0.42           &{\cellcolor[rgb]{0.8352941, 0.99607843, 0.6392157}}0.58           &{\cellcolor[rgb]{0.78039217, 0.9882353, 0.52156866}}0.53           &{\cellcolor[rgb]{0.9529412, 0.8392157, 0.8392157}}0.32           &{\cellcolor[rgb]{0.75686276, 0.9843137, 0.48235294}}0.42           &{\cellcolor[rgb]{0.95686275,1.0, 0.91764706}}0.28           \\
\textbf{C2}        &{\cellcolor[rgb]{0.91764706, 0.6901961, 0.68235296}}0.40           &{\cellcolor[rgb]{0.8627451, 0.99215686, 0.7019608}}0.42           &{\cellcolor[rgb]{0.972549, 0.92156863, 0.92156863}}0.31           &{\cellcolor[rgb]{0.972549, 0.92156863, 0.92156863}}0.60           &{\cellcolor[rgb]{0.9764706, 0.9137255, 0.9098039}}0.29           &{\cellcolor[rgb]{0.9254902, 0.7294118, 0.7176471}}0.30           &{\cellcolor[rgb]{0.9411765, 0.8, 0.7921569}}0.53           &{\cellcolor[rgb]{0.92941177, 0.73333335,0.7254902}}0.42           &{\cellcolor[rgb]{0.9137255, 0.7058824, 0.69803923}}0.28           &{\cellcolor[rgb]{0.96862745, 0.8980392, 0.8980392}}0.32          &{\cellcolor[rgb]{0.78431374, 0.9882353, 0.53333336}}0.32            \\
\textbf{C3}        &{\cellcolor[rgb]{0.7607843, 0.9882353, 0.49411765}}0.62           &{\cellcolor[rgb]{0.92941177, 0.73333335, 0.7254902}}0.33           &{\cellcolor[rgb]{0.972549, 0.92156863,0.92156863}}0.31           &{\cellcolor[rgb]{0.972549, 0.92156863, 0.92156863}}0.60           &{\cellcolor[rgb]{0.7607843, 0.9882353, 0.49411765}}0.38           &{\cellcolor[rgb]{0.9647059, 0.8627451, 0.85882354}}0.33           &{\cellcolor[rgb]{0.9411765, 0.99607843,0.88235295}}0.57           &{\cellcolor[rgb]{0.81960785,0.9882353,0.6039216}}0.52           &{\cellcolor[rgb]{0.9137255, 0.7058824, 0.69803923}}0.28           &{\cellcolor[rgb]{0.91764706, 0.70980394, 0.7019608}}0.27          &{\cellcolor[rgb]{0.9254902, 0.7372549, 0.7294118}}0.22            \\ 
\hline
\textbf{Avg. Acc.} & 0.46           & 0.40           &  0.32          & 0.61           & 0.30           & 0.36           &  0.57          & 0.50           &  0.33         & 0.33           &  0.27           \\
\textbf{SD}        &0.50            & 0.49           & 0.47           & 0.49           & 0.46           & 0.48           &  0.50          &  0.50          &  0.47          &  0.47          & 0.44            \\
\hline
\textbf{p-value}        &0.10            & 0.73            & 0.73           & 0.76       &0.58            & 0.51           &  0.94          & 0.57           &  0.17          & 0.37            & 0.64            \\
\bottomrule
\end{tabular}}
\caption{The accuracy of error handling for different types of errors under each condition.}
\label{tab:cond-errorType-accuracy}
\end{table}

Furthermore, we found that different error-handling mechanisms did not significantly influence the accuracy of SQL query error handling at various difficulty levels (Table~\ref{tab:cond-diff-acc}). These findings suggest that existing interaction mechanisms are not very effective for handling NL2SQL errors that state-of-the-art deep learning NL2SQL models make on complex datasets like Spider. We further discuss the reasons behind these results and their implications in the rest of Section~\ref{sec:findings} and Section~\ref{sec:discussion}.

\begin{table}[!htb]
\centering
\resizebox{0.33\linewidth}{!}{
\begin{tabular}{c|ccc} 
\toprule
                  & \multicolumn{3}{c}{\textbf{Difficulty levels 
                  }}   \\
                  & \textbf{Easy} & \textbf{Medium} & \textbf{Hard}  \\ 
\hline
\textbf{B1}       &{\cellcolor[rgb]{0.91764706, 0.70980394, 0.7019608}}0.64               &{\cellcolor[rgb]{0.91764706, 0.99607843, 0.8235294}}0.64                 &{\cellcolor[rgb]{0.91764706, 0.70980394, 0.7019608}}0.21                \\
\textbf{C1}       &{\cellcolor[rgb]{0.9882353, 0.9764706, 0.9764706}}0.71               &{\cellcolor[rgb]{0.91764706, 0.99607843, 0.8235294}}0.64                &{\cellcolor[rgb]{0.75686276, 0.9843137, 0.48235294}}0.36                \\
\textbf{C2}       &{\cellcolor[rgb]{0.75686276, 0.9843137, 0.48235294}}0.79               &{\cellcolor[rgb]{0.7490196, 0.9843137, 0.45490196}}0.71                 &{\cellcolor[rgb]{0.75686276, 0.9843137, 0.48235294}}0.36                \\
\textbf{C3}       &{\cellcolor[rgb]{0.75686276, 0.9843137, 0.48235294}}0.79               &{\cellcolor[rgb]{0.91764706, 0.69411767, 0.6862745}}0.50                 &{\cellcolor[rgb]{0.9882353,1.0, 0.98039216}}0.29                \\ 
\hline
\textbf{Avg. Acc} &0.73               &0.63                 &0.30                \\
\textbf{SD}       &0.45               &0.49                 &0.46                \\ 
\hline
\textbf{p-value}  &0.81               &0.71                 &0.83                \\
\bottomrule
\end{tabular}}
\caption{The error-handling of different difficulty levels under each condition}
\label{tab:cond-diff-acc}
\end{table}

\paragraph{\textbf{F2: The error handling mechanisms do not significantly impact the overall time of completion}} To study the impact of different error handling mechanisms on time usage, we analyzed the time of completion (ToC) of the query that was solved correctly by the participants. We used the same ANOVA test as applied in the previous analysis to test the mean difference among ToC using various error handling mechanisms (Table~\ref{tab:cond-acc-toc}), no significant significance was found among the groups ($p=0.52$).

Similarly, we analyzed the impact of different error-handling mechanisms on the selected error types. In general, the baseline method was more efficient in solving a task, while the conversational dialog system took more time compared with other methods. The results are shown in Table~\ref{tab:cond-errorType-toc}.

Additionally, the results of experiments on SQL queries of various levels of difficulty revealed differences among the error-handling mechanisms tested in the case of easy queries ($p=0.04$). Specifically, direct editing was found to be the fastest method when the query was easy, followed by the explanation and example-based approach (C1), the explanation-based visualization approach (C2), and the conversational dialog system (C3).

\begin{table}[!htb]
\centering
\resizebox{0.8\linewidth}{!}{
\begin{tabular}{c|ccccc|cccccc} 
\toprule
                   & \multicolumn{5}{c|}{\textbf{Syntactic types 
                   }}                  & \multicolumn{6}{c}{\textbf{Semantic types 
                   }}                                  \\
                   & \textbf{A} & \textbf{B} & \textbf{C} & \textbf{D} & \textbf{E} & \textbf{a} & \textbf{b} & \textbf{c} & \textbf{d} & \textbf{e} & \textbf{f}  \\ 
\hline
B1       &{\cellcolor[rgb]{0.91764706, 0.99607843, 0.81960785}}112.0                      &{\cellcolor[rgb]{0.77254903,0.9882353,0.50980395}}98.5                      &{\cellcolor[rgb]{0.83137256,0.99215686,0.63529414}}97.5                       & {\cellcolor[rgb]{0.9882353,0.9764706,0.9764706}}109.7                      &{\cellcolor[rgb]{0.99215686,1, 0.9882353}}109.6                       &{\cellcolor[rgb]{0.9137255,0.99607843,0.81960785}}104.0                       &{\cellcolor[rgb]{0.7137255,0.9882353,0.4}}93.8                      &{\cellcolor[rgb]{0.80784315,0.9882353,0.58431375}}116.0                      &{\cellcolor[rgb]{0.91764706,0.6901961,0.68235296}}130.0                      &{\cellcolor[rgb]{0.93333334,0.7607843,0.75686276}}121.7                    &{\cellcolor[rgb]{0.9882353,1,0.9764706}}103.0                       \\
C1       &{\cellcolor[rgb]{0.7372549,0.9882353,0.4509804}}103.3                      &{\cellcolor[rgb]{0.9764706,1.0,0.9647059}}103.8                      &{\cellcolor[rgb]{0.7294118,0.9843137,0.42352942}}91.1                       &{\cellcolor[rgb]{0.88235295,0.99215686,0.7411765}}104.6                     &{\cellcolor[rgb]{0.84313726,0.99215686,0.65882355}}101.7                       &{\cellcolor[rgb]{0.69803923,0.9882353,0.37254903}}96.1                       &{\cellcolor[rgb]{0.9019608,0.99215686,0.7882353}}103.9                      &{\cellcolor[rgb]{0.7176471,0.9843137,0.40392157}}110.1                      &{\cellcolor[rgb]{0.7372549,0.9882353,0.4509804}}107.0                      &{\cellcolor[rgb]{0.7294118,0.9882353,0.42745098}}97.9                       &{\cellcolor[rgb]{0.7137255,0.9882353,0.40392157}}83.4                        \\
C2       &{\cellcolor[rgb]{0.9882353,0.9607843,0.9607843}}117.9                      &{\cellcolor[rgb]{0.9490196,0.8235294,0.81960785}}108.2                      &{\cellcolor[rgb]{0.9098039,0.6862745,0.6784314}}86.2                       &{\cellcolor[rgb]{0.73333335,0.9882353,0.43137255}}96.8                       &{\cellcolor[rgb]{0.7254902,0.9843137,0.41568628}}93.0                        &{\cellcolor[rgb]{0.7882353,0.9882353,0.54509807}}99.5                       &{\cellcolor[rgb]{0.9411765,,0.7882353,0.7882353}}119.0                      &{\cellcolor[rgb]{0.99607843,0.9843137,0.9882353}}129.7                      &{\cellcolor[rgb]{0.78039217,0.9882353,0.5254902}}108.7                     &{\cellcolor[rgb]{0.8784314,0.99215686,0.73333335}}105.2                       &{\cellcolor[rgb]{0.88235295,0.99215686,0.7372549}}95.0                       \\
C3       &{\cellcolor[rgb]{0.91764706,0.7019608,0.69411767}}129.2                       &{\cellcolor[rgb]{0.92941177,0.7254902,0.72156864}}110.3                      &{\cellcolor[rgb]{0.91764706,0.7176471,0.7058824}}123.6                       &{\cellcolor[rgb]{0.9098039,0.6784314,0.67058825}}125.8                       &{\cellcolor[rgb]{0.9098039,0.6666667,0.65882355}}133.8                       &{\cellcolor[rgb]{0.9019608,0.63529414,0.61960787}}118.4                       &{\cellcolor[rgb]{0.9098039,0.654902,0.64705884}}125.0                      &{\cellcolor[rgb]{0.9098039,0.65882355,0.64705884}}148.6                       &{\cellcolor[rgb]{0.9882353,0.9764706,0.972549}}116.2                       &{\cellcolor[rgb]{0.9137255,0.6666667,0.65882355}}125.6                       &{\cellcolor[rgb]{0.9137255,0.65882355,0.654902}}125.0                       \\ 
\hline
Avg. ToC &115.6  &105.2  &99.6  &109.2  &109.5  &104.5  &110.4  &126.1  &115.5  &106.3  &101.6  \\
SD &33.2  &68.8  &48.5  &45.0  &73.1  &53.3  &37.5  &77.2  &73.6  &49.1  &59.3  \\
\hline
p-value &0.87  &0.99  &0.39  &0.60  &0.70  &0.82  &0.24  &0.71  &0.91  &0.17  &0.48  \\
\bottomrule
\end{tabular}}
\caption{The average ToC of different error types under each condition.}
\label{tab:cond-errorType-toc}
\end{table}

\begin{table}[!htb]
\centering
\resizebox{0.33\linewidth}{!}{
\begin{tabular}{c|ccc} 
\toprule
                  & \multicolumn{3}{c}{\textbf{Difficulty levels 
                  }}   \\
                  & \textbf{Easy*} & \textbf{Medium} & \textbf{Hard}  \\ 
\hline
\textbf{B1}       & {\cellcolor[rgb]{0.70980394,0.9843137,0.39607844}}31.8          &{\cellcolor[rgb]{0.7490196,0.9843137,0.4627451}}124.4           &{\cellcolor[rgb]{0.77254903,0.9843137,0.50980395}}133.6          \\
\textbf{C1}       & {\cellcolor[rgb]{0.92941177,0.99607843,0.85490197}}55.8          &{\cellcolor[rgb]{0.83137256,0.99607843,0.627451}}110.4           &{\cellcolor[rgb]{0.9372549,0.99607843,0.8666667}}154.2          \\
\textbf{C2}       & {\cellcolor[rgb]{0.9490196,0.83137256,0.83137256}}79.0          &{\cellcolor[rgb]{0.8352941,0.99607843,0.63529414}}110.5           &{\cellcolor[rgb]{0.90588236,0.6509804,0.6431373}}199.1          \\
\textbf{C3}       & {\cellcolor[rgb]{0.90588236,0.6509804,0.6431373}}95.7         &{\cellcolor[rgb]{0.91764706,0.69411767,0.6862745}}137.7           &{\cellcolor[rgb]{0.7294118,0.9843137,0.42352942}}125.3          \\ 
\hline
\textbf{Avg. ToC} & 65.6              & 120.7                & 153.1               \\
\textbf{SD}       & 50.9              & 96.2                & 97.7               \\ 
\hline
\textbf{p-value}  & 0.04          & 0.60            & 0.39           \\
\bottomrule
\end{tabular}}
\caption{The average ToC of different difficulty levels under each condition. *statistically significant difference ($p<0.05$)}
\label{tab:cond-diff-toc}
\end{table}

\paragraph{\textbf{F3: Users perform better on error types with fewer variants}} We analyzed the impact of error types on task accuracy and ToC, and reported the results in Table~\ref{tab:err-acc-duration}. The results revealed that among the syntactic error types, \texttt{A: WHERE errors} and \texttt{E: GROUP BY errors} had high accuracy, while for semantic error types, \texttt{d: Value error} and \texttt{e: Condition error} had high accuracy. As shown in the error taxonomy (Table~\ref{tab:taxonomy}), value errors occur only in the WHERE clauses, and those errors usually require fewer steps to fix and have little relationship with the other syntactic parts in an SQL query. Similarly, condition errors such as \texttt{wrong sorting directions} and \texttt{wrong boolean operator (AND, OR, etc.)} are relatively independent components in a query. The better user performance on those error types may indicate that users face challenges in handling \textit{semantically complicated} errors, such as joining tables and selecting columns from multiple tables, but are more successful in discovering and repairing error types where the error is more \textit{local} (i.e., with little interdependency with other parts of the query). This conclusion is also evidenced in the user interview, which we will analyze in the following section. 

\begin{table}[!htb]
\centering
\resizebox{0.9\linewidth}{!}{
\begin{tabular}{c|cc|cc} 
\toprule
\textbf{Syntactic types 
}                    & \textbf{Avg. Acc. ($\mu=0.53$)} & \textbf{SD  ($\mu=0.50$)} & \textbf{Avg. ToC ($\mu=128.8$) 
} & \textbf{SD ($\mu=91.4$)}  \\ 
\hline
\textbf{A}                                         &{\cellcolor[rgb]{0.75686276,0.9882353,0.47843137}}0.56                & {\cellcolor[rgb]{0.95686275,0.8509804,0.8509804}}0.51        &{\cellcolor[rgb]{0.9843137,0.94509804,0.94509804}}132.3              &{\cellcolor[rgb]{0.9137255,0.68235296,0.6784314}}105.5        \\
\textbf{B}                                         &{\cellcolor[rgb]{0.91764706,0.70980394,0.69803923}}0.48                & {\cellcolor[rgb]{0.972549,0.92156863,0.92156863}}0.50        &{\cellcolor[rgb]{0.9137255,0.68235296,0.6784314}}147.2              &{\cellcolor[rgb]{0.9843137,0.96862745,0.9647059}}90.4        \\
\textbf{C}                                         &{\cellcolor[rgb]{0.99607843,0.9882353,0.99215686}}0.53                & {\cellcolor[rgb]{0.95686275,0.8509804,0.8509804}}0.51        &{\cellcolor[rgb]{0.9843137,1,0.972549}}128.2              &\cellcolor[rgb]{0.78431374,0.9882353,0.5372549}75.6        \\
\textbf{D}                                         &{\cellcolor[rgb]{0.99607843,0.9882353,0.99215686}}0.53                & {\cellcolor[rgb]{0.8784314,0.99215686,0.7411765}}0.47        &{\cellcolor[rgb]{0.73333335,0.9882353,0.43137255}}111.5              &{\cellcolor[rgb]{0.972549,0.9098039,0.90588236}}55.5        \\
\textbf{E}                                         &{\cellcolor[rgb]{0.81960785,0.9882353,0.6039216}}0.55                & {\cellcolor[rgb]{0.95686275,0.8509804,0.8509804}}0.51        &{\cellcolor[rgb]{0.9372549,0.99607843,0.87058824}}125.1              &{\cellcolor[rgb]{0.7372549,0.9882353,0.43529412}}72.4         \\ 
\hline
\textbf{Semantic types 
} & \textbf{Avg. Acc. ($\mu=0.54$)}  & \textbf{SD ($\mu=0.50$)} & \textbf{Avg. ToC ($\mu=123.1$) (N=26)} & \textbf{SD ($\mu=80.53$)}  \\ 
\hline
\textbf{a}                                         &{\cellcolor[rgb]{0.9254902,0.7176471,0.70980394}}0.47                & {\cellcolor[rgb]{0.9254902,0.99607843,0.84313726}}0.49        &{\cellcolor[rgb]{0.99215686,0.9843137,0.9843137}}123.0              &{\cellcolor[rgb]{0.75686276,0.9882353,0.4745098}}67.4        \\
\textbf{b}                                         &{\cellcolor[rgb]{0.99607843,0.9882353,0.99215686}}0.54                & {\cellcolor[rgb]{0.9490196,0.8156863,0.8156863}}0.51        &{\cellcolor[rgb]{0.9882353,0.972549,0.9764706}}123.3              &{\cellcolor[rgb]{0.7647059,0.9882353,0.5058824}}68.2        \\
\textbf{c}                                         &{\cellcolor[rgb]{0.9529412,0.8352941,0.8352941}}0.50                &{\cellcolor[rgb]{0.9490196,0.8156863,0.8156863}}0.51        &{\cellcolor[rgb]{0.92941177,0.7647059,0.7607843}}128.7              &{\cellcolor[rgb]{0.78039217,0.9882353,0.52156866}}68.7        \\
\textbf{d}                                         &{\cellcolor[rgb]{0.7607843,0.9882353,0.49411765}}0.61                & {\cellcolor[rgb]{0.84705883,0.99215686,0.6784314}}0.47        &{\cellcolor[rgb]{0.8509804,0.99607843,0.6745098}}118.1              &{\cellcolor[rgb]{0.9137255,0.68235296,0.6784314}}96.5        \\
\textbf{e}                                         &{\cellcolor[rgb]{0.8,0.9843137,0.5647059}}0.60                &{\cellcolor[rgb]{0.9254902,0.99607843,0.84313726}}0.49        &{\cellcolor[rgb]{0.8,0.9882353,0.5764706}}116.7              & {\cellcolor[rgb]{0.9843137,0.95686275,0.9607843}}83.2        \\
\textbf{f}                                         &{\cellcolor[rgb]{0.9647059,0.8745098,0.8745098}}0.51                &{\cellcolor[rgb]{0.9490196,0.8156863,0.8156863}}0.51        &{\cellcolor[rgb]{0.9411765,0.7882353,0.7882353}}128.1              &{\cellcolor[rgb]{0.7490196,0.9843137,0.45490196}}66.8         \\
\bottomrule
\end{tabular}}
\caption{The average accuracy and ToC (in seconds) for different error types.}
\label{tab:err-acc-duration}
\end{table}

\paragraph{\textbf{F4: The explanation- and example-based methods are more useful for non-expert users}} When participants were asked to rate their preferences among the different interaction mechanisms (shown in Table~\ref{tab:preferences}), we found that the explanation- and example-based approach (C1) is the most preferred, while the explanation-based visualization approach (C2) was rated similarly to the baseline method (B1). In contrast, the conversational dialog system (C3) was generally rated as less useful than the others. \looseness=-1

\begin{table}[!htb]
\centering
\setlength{\extrarowheight}{0pt}
\addtolength{\extrarowheight}{\aboverulesep}
\addtolength{\extrarowheight}{\belowrulesep}
\setlength{\aboverulesep}{0pt}
\setlength{\belowrulesep}{0pt}
\resizebox{0.6\linewidth}{!}{
\begin{tabular}{c|cccc} 
\toprule
            & \textbf{Most useful}                  & \textbf{2nd most useful}               & \textbf{3rd most useful}              & \textbf{least useful}                 
\\ 
\hline
\textbf{B1} & {\cellcolor[rgb]{0.733,0.886,0.773}}7 & {\cellcolor[rgb]{0.831,0.929,0.729}}4  & {\cellcolor[rgb]{0.667,0.855,0.8}}9   & {\cellcolor[rgb]{0.765,0.898,0.757}}6  
\\
\textbf{C1} & {\cellcolor[rgb]{0.533,0.8,0.855}}13  & {\cellcolor[rgb]{0.635,0.843,0.816}}10 & {\cellcolor[rgb]{0.867,0.941,0.718}}3 & {\cellcolor[rgb]{0.965,0.984,0.675}}0 
\\
\textbf{C2} & {\cellcolor[rgb]{0.8,0.914,0.745}}5   & {\cellcolor[rgb]{0.702,0.871,0.788}}8  & {\cellcolor[rgb]{0.667,0.855,0.8}}9   & {\cellcolor[rgb]{0.831,0.929,0.729}}4 
\\
\textbf{C3} & {\cellcolor[rgb]{0.933,0.969,0.69}}1  & {\cellcolor[rgb]{0.831,0.929,0.729}}4  & {\cellcolor[rgb]{0.8,0.914,0.745}}5   & {\cellcolor[rgb]{0.435,0.757,0.898}}16 
\\
\bottomrule
\end{tabular}}
\caption{The participants' ranked preferences for different error handling mechanisms}
\label{tab:preferences}
\end{table}

We found that the user's level of expertise significantly impacts their adoption rate of different error-handling mechanisms. The adoption rate measures when a mechanism was available, and how likely that a user will use the mechanism (instead of just using the baseline method) to handle the error. We calculated the adoption rate for each condition (C1, C2, and C3) for different levels of expertise by dividing the number of SQL queries in which the participant used the provided error-handling mechanism by the total number of queries provided with the corresponding mechanism in the participant's study session. The result is shown in Table~\ref{tab:adoption}. 

\begin{table}[!htb]
\centering
\resizebox{0.55\linewidth}{!}{
\begin{tabular}{c|ccc}
\toprule
\textbf{Expertise levels}                  & \textbf{C1 ($\mu=0.74$)} & \textbf{C2 ($\mu=0.74$)} & \textbf{C3 ($\mu=0.41$)}  \\
\hline
\textbf{Expert}                      &  {\cellcolor[rgb]{0.9098039,0.6745098,0.6666667}}0.53         &  {\cellcolor[rgb]{0.90588236,0.6745098,0.67058825}}0.43         &  {\cellcolor[rgb]{0.99607843,0.99607843,0.9882353}}0.41          \\
\textbf{Intermediate}                &  {\cellcolor[rgb]{0.7607843,0.9882353,0.4862745}}0.84         &  {\cellcolor[rgb]{0.73333335,0.9882353,0.42745098}}0.90         &  {\cellcolor[rgb]{0.84705883,0.99215686,0.6784314}}0.44          \\
\textbf{Novice}                      &  {\cellcolor[rgb]{0.7294118,0.9882353,0.42352942}}0.86         &  {\cellcolor[rgb]{0.7529412,0.9843137,0.4745098}}0.88         &  {\cellcolor[rgb]{0.9490196,0.8156863,0.8156863}}0.38          \\
\bottomrule
\end{tabular}}
\caption{The adoption rate of each mechanism among different expertise levels}
\label{tab:adoption}
\end{table}

The primary factor contributing to the lower level of interest in using error handling mechanisms among expert participants under the experimental conditions was their ability to efficiently identify and repair errors independently. For example, P2 stated that \textit{``It (the step-by-step execution function in C1) is very redundant and time-consuming to break down the SQL queries and execute the sub-queries, since most errors can be found at first glance.''} Another reason why expert users were less interested in using the error handling mechanisms was that they were not confident in the intermediate results they provided. P3, for example, noted that \textit{``Though the chatbot is capable of revising the erroneous SQL queries, I found it sometimes gives an incorrect answer and provides no additional clues for me to validate the new query.''} Therefore, several expert participants chose to repair the original SQL query instead of validating and repairing the newly generated query.


The study also showed that the conversational dialog system (C3) was the least preferred mechanism among users at all levels of expertise. One reason for this is the relatively low accuracy of the model in recognizing user intents from the dialog and automatically repairing the errors in the query. For example, P3 stated that \textit{``Though it sometimes predicts the correct query, for most of the times, the prediction is still erroneous.''} In addition, the chatbot did not provide explanations for its suggestions, so users had to spend significant effort to validate and repair the newly generated SQL queries. Furthermore, while the chatbot allowed manual input from users to intervene in the prediction process, such as pointing out erroneous parts and providing correct answers, it often introduced new errors while predicting the SQL. As noted by P7: \textit{``In one example, when I asked the chatbot to change the column name that was in SELECT, it somehow changes the column in JOIN as well.''} As a result, many users quickly became frustrated after using it for a few SQL queries.

\begin{figure}[!htb]
\centering
\includegraphics[width=\linewidth]{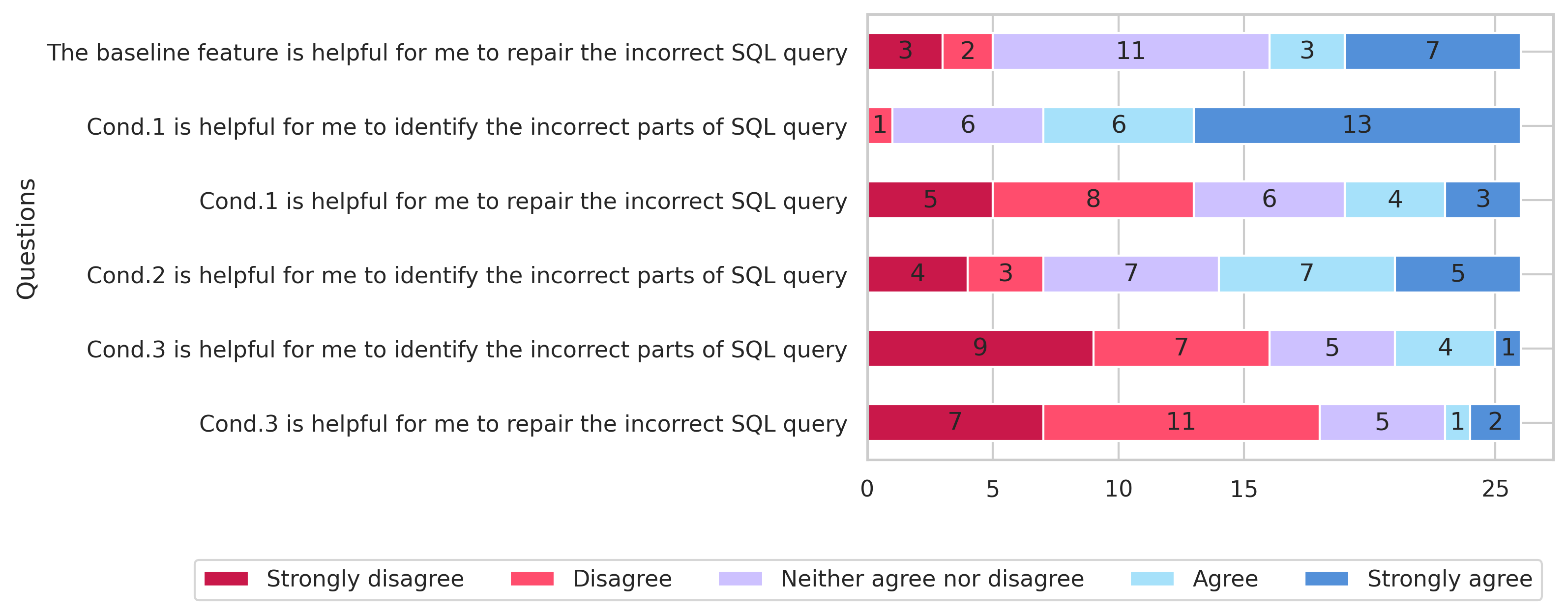}
\caption{The result of the post-study questionnaire}
\label{fig:post-study-questionnaire}
\end{figure}

\paragraph{\textbf{F5: The explanation- and example-based methods are more effective in helping users identify errors in the SQL query than in repairing errors}} In the post-study questionnaire, we asked participants to evaluate the usefulness of each condition in terms of its ability to help (1) identify and (2) repair errors, respectively (Fig.~\ref{fig:post-study-questionnaire}). The results indicate that most of the participants found C1 to be effective in identifying incorrect parts of the SQL query, while half of them thought it was not useful for repairing errors. Meanwhile, a notable proportion of participants (12 out of 26) affirmed C2's effectiveness in identifying the errors, but it was helpful for repairing the errors. In terms of C3, a significant number of participants (16 and 18) had a negative perception of its effectiveness in both identifying and repairing errors within the SQL query. \looseness=-1

Furthermore, we learned that the recursive natural language explanations might help reduce the understanding barrier for a long and syntactic-complicated SQL query. For example, P8 stated that \textit{``By looking at the shorter sentences first at the beginning, I could finally understand the meaning that the original long sentence were trying to convey.''} P17 also mentioned that: \textit{``Those shorter sentences usually did not have complex grammatical structures and confusing inter-table relationships, so that the problems were easier to be spotted.''} Additionally, executing the subquery and displaying the results were deemed helpful for localizing the erroneous parts in the original SQL query. For example, P23 stated: ``When I noticed that the retrieved result was empty, I realized that some problems should exist in the current query.'' In terms of C2, participants affirmed the effectiveness of graph-based SQL visualization in helping them better understand the relationship between the syntactical components of a query. The learning barrier of this approach was also the lowest among all experimental conditions: users could view the connections to a table by simply clicking the widget in the canvas. \looseness=-1

Then, we investigated why the participants were less satisfied with the effectiveness of repairing errors in a SQL query for C1. There were two main factors. First, the repair strategies supported by the error-handling mechanisms were limited. Specifically, participants could only \textit{replace} the incorrect parts with their correct substitutions using the drop-down menu of entity mappings, but for queries that require the addition, deletion, or reorganization of clauses, users had to manually edit the query. This limitation led to frustration among participants and ultimately resulted in them not prioritizing using this error-handling mechanism for future tasks. Second, the current approach provided little assistance for users in validating their edits. As a result, one participant stated that: ``I did not trust my own edits nor the suggested changes from the approach.'' (P20).

\section{Discussion and implications}
\label{sec:discussion}

\subsection{Improving NL2SQL model evaluations through the error taxonomy}

Currently, the evaluation of NL2SQL models emphasizes their accuracy from benchmark datasets. Though it is fair and effective in indicating the overall performance of the model, it fails in evaluating the model at a more fine-grained level, which impedes the development of error-handling strategies and the model's real-world application.

The error taxonomy we contributed helps us understand the types of syntactic and semantic errors that a particular NL2SQL model tends to make in addition to only the overall accuracy. This information can provide model developers with specific information to improve the model's robustness against certain error types, \ning{thereby developing more accurate NL2SQL models. Future work can focus on the technical solutions on how to design better NL2SQL models based on the taxonomy.}
Second, this taxonomy allowed us to make fine-grained comparisons between models beyond the accuracy metrics. By comparing the error distributions of different models, we can identify not only the relative advantages of individual models but also the common errors that current models are prone to make. 

Our work delves deep into the errors of the models, revealing that though model architectures and performances differ, they all exhibit high error rates in particular error types. On the other hand, some models, though have a relatively low overall accuracy, are capable of handling particular types of errors. Future work could focus on studying one of those high error-rate error types to increase the model performance.

\subsection{Design opportunities for NL2SQL error handling mechanisms}

The result of our empirical study suggests that existing error handling mechanisms do not perform as well on errors made by state-of-the-art deep-learning-based NL2SQL models on complex cross-domain tasks, despite the promising results reported in the respective evaluations of these mechanisms. We think the main reason could be that our study used a much more challenging dataset than what was used in prior studies. We used queries from Spider~\cite{yu-etal-2018-spider} (which is complex and cross-domain) that the state-of-the-art of NL2SQL models (instead of the earlier NL2SQL models, which would start to make errors on simpler SQL queries) failed on. The dataset used in our study more accurately represents realistic error scenarios that users encounter in natural language data queries. Here, we identified several design opportunities for more effective NL2SQL error-handling mechanisms.


\subsubsection{Enabling effective mixed-initiative collaboration between users and error handling tools}
Our findings indicate that the current error-handling tools for NL2SQL models do not provide sufficient feedback to users when they attempt to modify SQL queries. While existing error-handling mechanisms, such as the conversational dialog approach (C3), have focused on predicting correct modifications using static forms of user input, they have not adequately addressed the need for mechanisms to elicit useful human feedback to guide model prediction. For example, in C3, users provide input in the form of multiple-choice options for the recommended locations of potential errors, which was considered confusing and not useful by some participants, particularly when ``none of the recommended options made sense'' (P15) or ``the errors existed in multiple places and cannot be fixed by only selecting one answer'' (P24). Therefore, we suggest that future work should focus on the development of effective mixed initiative mechanisms that allow both users and error-handling tools to develop a mutual understanding of the model's current state of understanding and the user's intent.\looseness=-1 

\subsubsection{Implementing interactive approaches based on attention alignment.}
Our study and analysis on attention have proved the correlation between erroneous queries and attention misalignment, suggesting that a potential way of designing an error-handling mechanism is to enable users to correct the misalignment.
However, a majority of existing work focuses on automated attention alignment in the decoding process of an NL2SQL model without involving humans.
For example, some work~\cite{attention1, attention5, attention6} uses an external model trained on a dataset of human attention to adjust the model attention, while some works may use a statistical~\cite{attention4} approach.
Nevertheless, aligning attention automatically has limitations such as requiring the design of a model-specific alignment mechanism and the preparation of a dataset of human attention, which can not be generalized efficiently.

Our study of model attention provides a supportive theory for designing interactive approaches for NL2SQL models. In fact, existing interactive mechanisms can also be viewed as the implementation of attention alignment between the user and the model. For instance, MISP~\cite{misp} asks clarification questions to users when uncertain about a generated token. This QA procedure aims to force the model's attention to align with the user's attention when the model's attention is very likely to be deviated. 
Our study also highlights the possibility of designing an attention-based error-handling mechanism. For example, the interactive approach can enable users to comprehend and directly manipulate the model's attention.
Various approaches may employ different mechanisms at different layers of attention.
But as long as humans are included in the loop, the model has an opportunity to align its attention with the human's attention through the model's explanation and user feedback.

\subsubsection{Comprehending the generated queries and inspecting how queries operate on data complement each other}
The results of the study suggested that, to support effective NL2SQL error handling for users, it is important to help users (1) interpret the meaning of the generated SQL query, untangle its structures, and explain how it corresponds to the user's NL query; and (2) inspect the behaviors of the query on example data and examine whether they match user intents and expectations. The two parts are interdependent on each other. In practice, the user's preferences for these different approaches may vary depending on their expertise. For example, in our study, non-users and novice SQL users appreciated the explanation-based visualization mechanism (in SQLVis~\cite{SQLVis}) and the NL explanations in step-by-step execution of the generated queries (in DIY~\cite{narechania2021diy}), because these mechanisms lower the barrier to understanding the generated SQL queries for users who are unfamiliar with SQL syntax and structures. This preference was also reflected in their use of different mechanisms in the study. Experienced SQL users, on the contrary, did not use mechanisms for explaining the meanings of the generated SQL queries as often. However, they found the entity mapping feature and the example tables (in DIY) useful for discovering NL2SQL errors.



\subsubsection{Opportunities for adaptive strategies}
Lastly, the results of our user study suggest that the most effective error-handling strategy to use depends on many factors such as user expertise, query type, and possible error types. For example, expert users may require less sense-making strategies (e.g., step-by-step NL explanation), while they may expect an intuitive execution result preview or an efficient validation of the updated answer. In contrast, intermediate or novice users may need more mixed-initiative guides to facilitate error discovery and repair. Meanwhile, as discussed in Section~\ref{sec:findings}, the length, syntactical components, and potential error types of a query would result in different barriers to users when repairing errors. For example, for queries with more complicated syntactical structures, a visualization-based approach might be useful to reduce the barrier to understanding the structure of the query. Therefore, we recommend that future work in this area consider the development of adaptive error-handling strategies. An effective NL2SQL system could adapt its interface features, models, and interaction strategies according to the use case and context. Specifically, it could consider the semantic and syntactic characteristics of the query, whether the error is local (i.e., on a specific entity in the query) or global (i.e., regarding the overall query structure), and the user's preferences and level of expertise.\looseness=-1

\section{Limitations and Future Work}
\label{sec:limitations}
The current study has several limitations. First, the total number is unbalanced for each error type (as shown in Section~\ref{sec:error_distribution}), which may cause bias in the study of error-handling mechanisms in Section~\ref{sec:study_error_handling_mechanisms}. Despite the fact that Spider is already a large-scale dataset, there were only a small number of example errors in some rare error types. Therefore, we have to exclude these error types in our analysis. The problem could be addressed by conducting larger-scale user studies with more participants and erroneous query data.

Second, despite that we reproduced four representative NL2SQL models based on the model architecture, it is hard to cover all due to the lack of open-source implementation or the engineering challenges in adapting them to our analysis pipeline. In addition, all the models used in our study are ``black-box'' models that do not provide much transparency into the process, which limits the selection of error-handling mechanisms. Interactive models~\cite{10.1145/3209978.3210013,Elgohary2021NLEDITCS}, on the other hand, provide the transparency that could allow additional error handling mechanisms, such as modifying the intermediate results of the model predictions. In future work, we will expand the scope of our research to include additional types of representative NL2SQL models.


Third, in the cause analysis study, we only explored one of the possible causes --- attention misalignment, a more comprehensive analysis could be conducted to build up the factors that contribute to the model's performance. Additionally, we only tested one model in our study due to the computational resource and time limit. In future work, it could be valuable to explore recently emerged large open-sourced foundational models such as LlaMa-2~\cite{llama}.

Lastly, while the example SQL queries were real erroneous queries made by NL2SQL models on realistic databases and natural language queries, the setting of our study is still quite artificial, lacking the real-world task context in the actual usage scenarios of NL2SQL systems. In the future, it will be useful to study user error handling behaviors through a field study to better understand the impact of task-specific contexts on user behavior and the effectiveness of user handling of NL2SQL errors.

\section{Conclusion}
In this paper, we i) presented an empirical study to understand the error types in the SQL query generated by NL2SQL models; ii) explored a possible cause of model errors by measuring the alignment of model-user attention on the NL query, and iii) conducted a controlled user experiment with 26 participants to measure the effectiveness and efficiency of representative NL2SQL error handling mechanisms. The error taxonomy summarized 48 types of errors and revealed their distribution characteristics. Our study also demonstrated a strong correlation between the cause of model errors and the misalignment of attention between humans and models. The results of the user experiment revealed challenges and limitations of existing NL2SQL error handling mechanisms on errors made by state-of-the-art deep-learning-based NL2SQL models on complex cross-domain tasks. Based on the results, we identified several research opportunities and design implications for more effective and efficient mechanisms for users to discover and repair errors in natural language database queries.

\begin{acks}
This research was supported in part by an AnalytiXIN Faculty Fellowship, an NVIDIA Academic Hardware Grant, a Google Cloud Research Credit Award, NSF grant CCF-2211428, and NSF grant ITE-2333736. Any opinions, findings, or recommendations expressed here are those of the authors and do not necessarily reflect the views of the sponsors. 
\end{acks}

\bibliographystyle{ACM-Reference-Format}
\bibliography{references}

\end{document}